\documentclass[12pt]{article}
\usepackage[cmtip,arrow]{xy}\usepackage{pb-diagram,pb-xy}%

\input xy
\xyoption{all}
\input xy
\xyoption{all}
\usepackage{heck}
\usepackage{cite}
\usepackage{graphicx}
\usepackage{makeidx}
\usepackage{multicol}
\usepackage{amsfonts}
\usepackage{mathrsfs}
\usepackage{amssymb}
\usepackage{amsmath}%

\providecommand{\U}[1]{\protect\rule{.1in}{.1in}}
\numberwithin{equation}{section}

\newcommand{\bea}{\begin{eqnarray}}
\newcommand{\eea}{\end{eqnarray}}
\newcommand{\be}{\begin{equation}}
\newcommand{\ee}{\end{equation}}

\newcommand{\bem}{\begin{pmatrix}}
\newcommand{\eem}{\end{pmatrix}}

\def\a{\alpha}

\def\U{\Upsilon}

\def\cc{{\cal C}}

\def\cn{{\cal N}}

\def\cp{{\cal P}}

\def\cw{{\cal W}}

\def \Z {{\mathbb Z}}
\def \C {{\mathbb C}}
\def \R {{\mathbb R}}

\xyoption{arc}

\date{July, 2012}

\institution{SISSA}{\centerline{SISSA, via Bonomea 265, I--34100 Trieste, ITALY}}

\title{4d $\cn=2$ Gauge Theories and Quivers:\\
the Non--Simply Laced Case}
\authors{Sergio Cecotti\footnote{e-mail: {\tt cecotti@sissa.it}} and Michele Del Zotto\footnote{e-mail: {\tt eledelz@gmail.com}}%
}

\abstract{We construct the BPS quivers with superpotential for the 4d $\cn=2$ gauge theories with \emph{non}--simply laced Lie groups ($B_n$, $C_n$, $F_4$ and $G_2$). The construction is inspired by the BIKMSV geometric engineering of these gauge groups as \emph{non--split} singular elliptic fibrations. From the categorical viewpoint of {\tt arXiv:1203.6743}, the fibration of the light category $\mathscr{L}(\mathfrak{g})$ over the (degenerate) Gaiotto curve has a monodromy given by the action of the outer automorphism of the corresponding \textit{unfolded} Lie algebra.

In view of the Katz--Vafa `matter from geometry' mechanism, the monodromic idea may be extended to the construction of $(Q,\cw)$ for SYM coupled to higher matter representations. This is done through a construction we call \emph{specialization}.    
}

\begin{document}

\maketitle

\tableofcontents
\newpage

\section{Introduction}

In the last year there has been significant progress in the computation of BPS spectra of 4d $\cn=2$ supersymmetric gauge theories using the Representation Theory of quivers with superpotentials \cite{CNV,CV11,ACCERV1,ACCERV2,arnold, cattoy, half} (for previous work see \cite{Denef00,DM07,DM,Dia99,DFR1,DFR2,FM00,Fiol,Denef,FHHI02,FHH00,Feng:2001xr,HK05,FHKVW05,Feng:2005gw}; for other approches to the $\cn=2$ BPS spectra see \cite{Shapere:1999xr,Gaiotto,GMN09,GMN10,GMN11,GMN12}). In the above references the discussion was always limited to theories whose gauge group $G$ is a product of Lie groups of type $ADE$. This limitation arose from the lack of a construction of the quivers with superpotential for the $\cn=2$ models with gauge group of type $BCFG$. The goal of the present paper is to fill this gap.\smallskip

The existence of a quiver description of the BPS spectrum for $\cn=2$ SYM with $ADE$ gauge groups follows \cite{CNV} from their geometric engineering in string theory as fibrations of $ADE$ singularities. From this point of view, the strategy to extend the quiver description to the non--simply laced gauge groups appears obvious:  in ref.\!\cite{BIKMSV} BIKMSV engineered  all non--simply laced gauge groups by considering \emph{non--split} singular elliptic fibrations (for a related construction see \cite{vafageo}). Roughly speaking, the dichotomy \textit{simply--laced \emph{vs.}\! non--simply--laced} corresponds geometrically to the dichotomy  \textit{split \emph{vs.}\! non--split}.  Thus one is naturally led to the question: `Is there a notion of split/non--split in the Representation Theory of a quiver with superpotential, $(Q,\cw)$, which corresponds to the BIKMSV geometric notion?' or, more directly, --- `Which $(Q,\cw)$'s describe a non--split situation?'
\smallskip

The general (and rather abstract) theory developed in \cite{cattoy} has a simple answer to this question (see equation \eqref{uuuwer} below). Knowing that answer, one may write down the $(Q,\cw)$'s for all $BCFG$ pure SYM in a matter of few minutes. The resulting pairs $(Q,\cw)$ are guaranteed by (geometric) construction to have the physically right BPS spectrum in the weak coupling limit $g_\mathrm{YM}\rightarrow 0$; in particular, in that limit the only stable particles with zero magnetic charge are vector--multiplets making precisely one copy of the adjoint of the gauge group $G$. 
\smallskip

In the same vein, we take inspiration from the Katz--Vafa `matter from geometry' construction \cite{matterfromgeo} to get the pairs $(Q,\cw)$ for SYM coupled to matter. This allows to determine $(Q,\cw)$ for matter representations more general than those discussed in refs.\!\cite{ACCERV2,cattoy}. We call this method \emph{specialization} of the quiver (with superpotential). Putting the two geometric constructions --- monodromy and specialization --- together, leads to the pairs $(Q,\cw)$ for  $BCFG$ SYM coupled to matter.
\smallskip

The present paper is organized as follows. In section 2, following \cite{BIKMSV,vafageo}, we describe the geometrical ideas on which our construction is based. In section 3 we determine the relevant quivers in two different ways: by relating them to certain integrable statistical systems, called $Q$--systems \cite{difrancesco1,difrancesco2,difrancesco3}, as well as by using the Dirac integrality condition introduced in \cite{cattoy,half}. In section 4 we review the notion of the light subcategory corresponding to the set of `perturbative' states \cite{cattoy}. Section 5 is the heart of the paper: there it is shown how a light category with the BIKMSV monodronic  properties arises naturally and uniquely from the special properties of the non--simply laced Dynkin graphs; the light BPS spectrum is computed and found to agree with the physically expected one. In section 6 we write down the superpotentials $\cw$ for all non--simply laced pure SYM theories. In section 7 we sketch other possible applications of the monodromic construction. In section 8 we present a preliminary discussion of the quivers for 
$BCFG$ SQCD with various matter content.  Technical aspects and detailed computations are confined in the appendices.

\section{Geometric preliminaries}\label{sec:geometric}

We refer to \cite{CV11,ACCERV1,ACCERV2,cattoy,half} for general background on the correspondence between the Representation Theory of quivers with superpotential and the BPS spectra of 4d $\cn=2$ QFTs, including the notion of \emph{stability} of a representation and the general properties of the stable representations.
\bigskip 

The description of the BPS sector of a 4d $\cn=2$ gauge theory in terms of stable representations of a quiver $Q$, bounded by the Jacobian relations $\partial\cw=0$ arising from a superpotential $\cw$, originates from the geometric construction of the gauge theory from string theory or $M$/$F$--theory. In this context the non--Abelian gauge groups arise from fibrations of $ADE$ singularities \cite{BIKMSV};  \textit{non}--simply laced gauge groups may also arise provided the fibration is \emph{non--split},  as explained in \cite{BIKMSV}.
This last case corresponds to the Slodowy generalization of the classical McKay correspondence to the non--simply laced case \cite{COXX}.\medskip

\begin{figure}
\begin{center}
\begin{gather*}
\begin{gathered}
\xymatrix{ &  &&& \bullet \ar@{-}[dl]\ar@{..>}@/^1.5pc/[dd]\\
\bullet \ar@{-}[r] &\cdots \ar@{-}[r] &\bullet \ar@{-}[r]& \bullet\\
&&&& \bullet \ar@{-}[ul]}\end{gathered}
\qquad  \longrightarrow\ 
\begin{gathered}
\xymatrix{ 
\bullet \ar@{-}[r] &\cdots \ar@{-}[r] &\bullet \ar@{-}[r]& \bullet\ar@{=>}[r]&\bullet}\end{gathered}\\
\phantom{mmmm}\\
\begin{gathered}
\xymatrix{& \bullet \ar@{-}[r] &\cdots \ar@{-}[r]& \bullet \ar@{-}[r] &\bullet\ar@{..>}@/^1.5pc/[dd]\\
 \bullet \ar@{-}[ur]\ar@{-}[dr] \\
&\bullet \ar@{-}[r] &\cdots \ar@{-}[r]& \bullet \ar@{-}[r] &\bullet}\end{gathered}
\qquad  \longrightarrow\ 
\begin{gathered}
\xymatrix{ 
\bullet \ar@{=>}[r] &\bullet \ar@{-}[r] &\cdots \ar@{-}[r] & \bullet\ar@{-}[r]&\bullet}\end{gathered}\\
\phantom{mmmm}\\
\begin{gathered}
\xymatrix{& \bullet\ar@{-}[dl]  \ar@{..>}@/^1pc/[d]\\
 \bullet \ar@{-}[r] &\bullet \\
&\bullet \ar@{-}[ul] \ar@{..>}@/_1pc/[u]}\end{gathered}
\qquad  \longrightarrow\ 
\begin{gathered}
\xymatrix{ 
\bullet \ar@3{->}[r] &\bullet }\end{gathered}\\
\phantom{mmmm}\\
\begin{gathered}
\xymatrix{&& \bullet \ar@{-}[r] &\bullet\ar@{..>}@/^1.5pc/[dd]\\
\bullet \ar@{-}[r]& \bullet \ar@{-}[ur]\ar@{-}[dr] \\
&&\bullet \ar@{-}[r] &\bullet}\end{gathered}
\qquad  \longrightarrow\ 
\begin{gathered}
\xymatrix{ 
\bullet \ar@{-}[r] &\bullet \ar@{=>}[r] & \bullet\ar@{-}[r]&\bullet}\end{gathered}
\end{gather*}
\caption{Dynkin diagrams foldings. The graphs on the left are called the `parent graph'  $G_\text{parent}$ of the graphs $G$ on the right.}
\label{fig;foldin}
\end{center}
\end{figure}
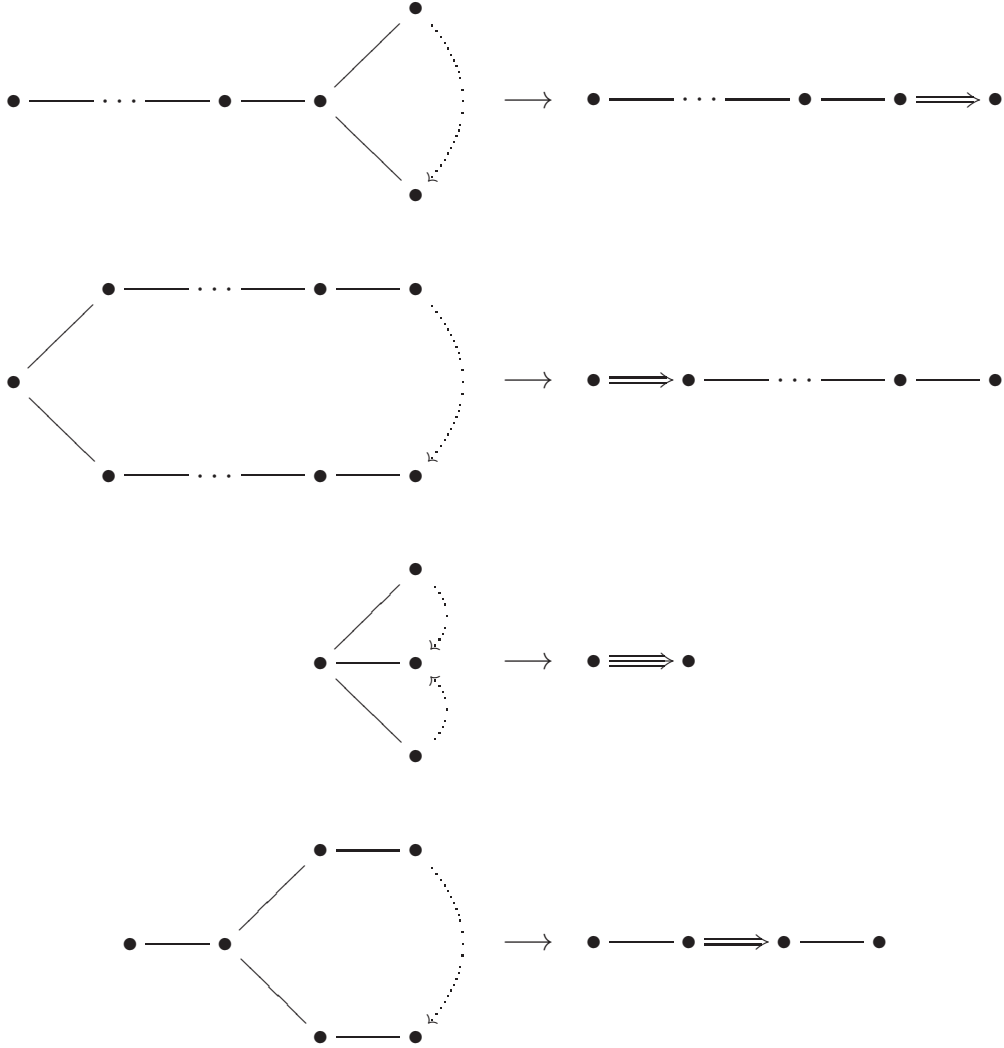

It is well--known \cite{COXX,bump} that the Dynkin graph of a non--simply laced Lie group $G$ arises by folding a parent  simply--laced Dynkin graph $G_\text{parent}$ along an automorphism. Specifically, the $G_\text{parent}\rightarrow G$ foldings are
\begin{gather}
D_{n+1}\longrightarrow B_n\\
A_{2n-1}\longrightarrow C_n\\
D_{4}\longrightarrow G_2\\
E_{6}\longrightarrow F_4,
\end{gather}
see figure \ref{fig;foldin}. In all cases the folding corresponds to a $\Z_2$ automorphism of the parent graph, except for the $G_2$ case, where we may take it to be $\Z_3$. To each node of the folded Dynkin diagrams in fig.\,\ref{fig;foldin} there is attached an integer $d_i$, namely the number of nodes of the parent graph which were folded into it. This number corresponds to one--half the length--square\footnote{The length of the short (long) co--roots (resp.\! roots) is normalized to 2.} of the corresponding simple co--root $\alpha_i^\vee$
\begin{equation}\label{defdi}
d_i= \frac{1}{2}(\alpha_i^\vee,\alpha_i^\vee)\equiv \frac{2}{(\alpha_i,\alpha_i)}\qquad i=1,2,\dots, r.
\end{equation}
\smallskip

The BIKMSV \cite{BIKMSV} construction of a non--simply laced gauge group is based on a $G_\text{parent}$ fibration having a non--trivial monodromy given by the action of the outer automorphism group $\Z_2$ ($\Z_3$) of the parent Lie algebra induced by the Dynkin diagram symmetry associated to the given folding (fig.\,\ref{fig;foldin}).

This monodronic construction may be imported in the context of Gaiotto engineering \cite{Gaiotto} of class--$S$ 4d $\cn=2$ gauge theories from  
the 6d $(2,0)$ theory of type $\mathfrak{g}=ADE$ as follows. In the Gaiotto framework, a weakly coupled SYM sector with simply--laced  gauge group $G_\mathrm{s.l.}=\exp[\mathfrak{g}]$ is realized by considering  a long thin tube in the UV surface $\cc$
\cite{Gaiotto}. If the tube is very long, we may work locally in its bulk, and replace the surface $\cc$ by an infinite cylinder $\C^\times$; this corresponds to ignoring the parts of the surface attached at the infinite ends of the tube which act as `matter', whose global symmetry is gauged by the given SYM sector. Neglecting the `matter' is physically justified, since a very long tube corresponds to a very small YM coupling, and then the `matter' asymptotically decouples.

Nothing prevents us from using a \emph{twisted} version of this construction parallel to the BIKMSV \emph{non--split} elliptic fibrations. (For a related construction in a slightly different setting see \cite{tackk}). Indeed, considering the $(2,0)$ 6d theory of type, respectively, $$\mathfrak{g}=D_{n+1}, A_{2n-1}, E_6, D_4$$ on such an infinite cylinder, $\C^\times$, times $\R^{1,2}\times S^1$, and reducing first on the $S^1$ factor, we are led \cite{GMN09} to solve the Hitchin equations for a $\mathfrak{g}\otimes K_\cc$ valued Higgs field $\varphi$ and the $\mathfrak{g}$--connection $A$ satisfying a twisted periodic boundary conditions. Let $\theta=\arg z$ be the circle coordinate on $\C^\times \simeq S^1\times \R$; we require
\begin{equation}\label{geotwisting}
\Big(A(\theta+2\pi,r),\;\varphi(\theta+2\pi,r)\Big)=\Big(\omega\cdot A(\theta,r),\;\omega\cdot\varphi(\theta,r)\Big),
\end{equation}
where $\omega$ is a non--trivial outer automorphism of $\mathfrak{g}$. Of course, this twisting is consistent only if we also twist the adjoint fields $A_\mu(x;\theta,r)$, $\psi(x;\theta,r)$ and $\Phi(x;\theta,r)$ which contain the 4d $\mathfrak{g}$--valued vector--multiplet. Then, from the 4d perspective, this `monodromic' geometry corresponds to a weakly coupled $\cn=2$ SYM theory with the non--simply laced gauge group $G$ whose Dynkin graph is obtained by folding the Dynkin diagram of $\mathfrak{g}$ along the cyclic group $\{\omega^k\}$. 

  \smallskip
  
To make contact between this geometrical construction and the Representation Theory of quivers, we have to recall the categorical interpretation of the degeneration limit of the Gaiotto curve presented in ref.\!\cite{cattoy}. If the gauge group $G=\exp \mathfrak{g}$ is simple and simply--laced, the SYM light category\footnote{\ The main properties of the light category $\mathscr{L}$ are reviewed in section 4 below.} $\mathscr{L}$ has the form\footnote{\ Given two Krull-Schimdt additive $\C$--categories, $\mathscr{A},\mathscr{B}$, we write $\mathscr{A}\vee \mathscr{B}$ for the category whose objects are the directs sums of objects of $\mathscr{A}$ and $\mathscr{B}$. The equation in the text then says that an object in $\mathscr{L}$ may be written as a (finite) direct sum $X_{\lambda_1}\oplus \cdots\oplus X_{\lambda_\ell}$ with $X_{\lambda_s}\in\mathscr{L}_{\lambda_s}$. } \cite{cattoy}
\begin{equation}\label{symdec}
\mathscr{L}=\bigvee_{\lambda \in \mathbb{P}^1}\mathscr{L}_\lambda,
\end{equation}
where $\mathscr{L}_\lambda$ are Abelian categories, (independent of $\lambda$ up to equivalence). The bricks\footnote{\ A representation $X$ is called a \emph{brick} iff $\mathrm{End}\,X=\C$. All stable representations are, in particular, bricks.} of $\mathscr{L}_\lambda$ are rigid and their images in the lattice $K_0(\mathscr{L})$ are positive roots of $\mathfrak{g}$ \cite{cattoy}. Physically, these two properties say that the BPs states at weak coupling with zero magnetic charge are vector--multiplets in the adjoint of the gauge group \cite{cattoy}. 
Cutting away the ends of the tube, corresponds to omitting the two fibers over the poles in the \textsc{rhs} of eqn.\eqref{symdec}, that is, to considering the `open' subcategory 
\begin{equation}\label{symdec2}
\bigvee_{\lambda \in \C^\times}\mathscr{L}_\lambda.
\end{equation}

An outer automorphism $\omega$ of $\mathfrak{g}$ induces an autoequivalence $\mathscr{G}$ in each category $\mathscr{L}_\lambda$. The categorical analogue of the geometric twisting
\eqref{geotwisting}  is then
\begin{equation}\label{uuuwer}
\mathscr{L}_{e^{2\pi i}\lambda}= \mathscr{G}\,\mathscr{L}_\lambda.
\end{equation}
In other words: there is an equivalence of categories of the form
\begin{equation}
\gamma\colon\ \mathscr{L}(G)_\lambda\longrightarrow \mathscr{L}(G_\mathrm{par})_\lambda
\end{equation}
which associates to each object of the SYM light category for the non--simply laced gauge group $G$, $\mathscr{L}(G)_\lambda$, an object of the SYM light category with the simply--laced gauge group $G_\mathrm{par}$, $\mathscr{L}(G_\mathrm{par})_\lambda$. $\gamma$ preserves all Hom spaces (in particular, it preserves the End rings).
However, the equivalence $\gamma$ is \emph{not} canonical since it depends on the choice of a root of the algebraic equation
\begin{equation}
x^n=\lambda,\qquad n\equiv \text{order}\;\omega.
\end{equation}
Changing the root will replace the equivalence by making 
\begin{equation}\gamma\rightarrow \gamma\circ\mathscr{G}^k,\end{equation}
for some $k\in\Z$.

\section{$Q$--systems and Dirac integrality conditions}\label{sec;dirac}
We have very natural candidates for the quiver of 
 all pure SYM with $G$ non--simply laced. Indeed, for each finite--dimensional simple Lie algebra $\mathfrak{g}$ there is a certain integrable statistical system --- known as the $Q$--system of type  $\mathfrak{g}$  \cite{difrancesco1,difrancesco2,difrancesco3} --- which is defined in terms of the cluster algebra associated to a certain mutation--class of quivers depending on $\mathfrak{g}$. 
For $\mathfrak{g}$ simply--laced, the $Q$--sysyem quivers coincide with those for pure $\mathfrak{g}$ SYM as described in refs.\!\cite{CNV,ACCERV2,cattoy}; it is natural to expect that the identification of the pure SYM quivers with the $Q$--system ones remains true also for $\mathfrak{g}$ \emph{non}--simply laced. Indeed, the (classical) $Q$--systems lead to recursion relations for characters of quantum groups \cite{difrancesco1} which, in turn, are crucial for the physical interpretation of $\mathsf{rep}(Q,\cw)$ as the space of BPS of $\cn=2$ SYM with the given gauge group.   
\smallskip

There is a general \emph{first principle} strategy to find the quiver of a $\cn=2$ gauge theory (if it exists at all), namely the Dirac integrality conditions \cite{cattoy,half}.  As we are going to show, this method confirms that the $Q$--systems quivers are the correct ones for pure SYM even in the non--simply laced case.\smallskip

Let $C$ be the Cartan matrix of $\mathfrak{g}$; then the exchange matrix of the (standard representative) quiver of the associated $Q$--system is the $2r\times 2r$ skew--symmetric matrix\footnote{\ In the conventions of \cite{difrancesco1}, this is the exchange matrix of the \emph{opposite} quiver $Q^\mathrm{op}$.}
\begin{equation}\label{exchange}
B=
\left(\begin{array}{c|c}
C-C^t & C^t\\\hline
-C & 0
\end{array}\right)
\end{equation}
In other words, the quiver for $\mathfrak{g}$ SYM is obtained by drawing $r=\mathrm{rank}\,\mathfrak{g}$ vertical Kronecker subquivers $\upuparrows$ and then connecting the sink of the $i$--th Kronecker to the source of the $j$--th one by $|C_{ij}|$ arrows. In addition, one draws 
$|C_{ij}-C_{ji}|$ extra `diagonal' arrows in such a way that the net number of arrows connecting any two vertical Kronecker subquivers is zero. The \emph{net} number of arrows entering (leaving) the source (resp. the sink) of the $i$--th Kronecker subquiver $\mathbf{Kr}_i$ is always equal to the integer $d_i$ defined in eqn.\eqref{defdi}.

 The extra `diagonal' arrows guarantee that the indecomposable representations $X_{\alpha_i}$ having support in the $i$--th vertical Kronecker subquiver $\mathbf{Kr}_i$ and dimension vector equal to the imaginary root $\delta_i$ of the $\widehat{A}_1$ affine  Dynkin graph underlying $\mathbf{Kr}_i$,
 are \emph{mutually local}. This last property is crucial for the physical consistency, since the representation $X_{\alpha_i}$ corresponds to the $\alpha_i$ simple--root $W$--boson, and the $W$--bosons must be mutually--local. 
 
 The charges of the $\alpha_i$ $W$--boson with respect to the maximal torus $U(1)^r\subset G$ are given by 
 \begin{equation}\label{polarynng}
 q_j(\alpha_i)\equiv \alpha_i(\alpha^\vee_j)\equiv \frac{2\, (\alpha_i,\alpha_j)}{(\alpha_j,\alpha_j)}\equiv C_{ij}.
 \end{equation}
 Let $\Gamma$ be the charge lattice of the 4d theory, which is identified with the dimension lattice of $\mathsf{rep}(Q,\cw)$ 
 \cite{CV11,ACCERV1,ACCERV2}.
 From eqn.\eqref{polarynng} we see that the vector in $\Gamma\otimes \mathbb{Q}$ corresponding to the unit $q_j$ charge is
 \begin{equation}
 \mathfrak{q}_j= (C^t)^{-1}_{ji}\:\dim X_{\alpha_i}.
 \end{equation}
 The magnetic weights $m_j(X)$ of the representation $X$ are then defined trough its Dirac electromagnetic pairing  with the unit electric vectors $\mathfrak{q}_j$ \cite{cattoy,half}
 \begin{equation}\label{magch1}
m_j(X)=- \langle\,\mathfrak{q}_j, \dim X\,\rangle_\text{Dirac}\equiv -(C^t)^{-1}_{ji}\, \langle\, \mathfrak{q}_j, \dim X\,\rangle_\text{Dirac} 
 \end{equation}
 \smallskip
 In the basis \eqref{exchange}, the charge vector of the $W$--boson associated to the $\alpha_i$ simple root of $\mathfrak{g}$  is then
\begin{equation}
 \dim X_{\alpha_i}\equiv \delta_i=  \Big(\ e_i\ \Big|\ e_i\ \Big)
\end{equation}
where $e_i$ stands for the row $r$--vector with $1$ in position $i$ and zero elsewhere. Hence 
\begin{equation}\label{didi1}
\langle\,\delta_i, \dim X\,\rangle_\text{Dirac}= \delta_i\,B\, (\dim X)^t =- (C^t\,\dim X)_{i,1}+(C^t\,\dim X)_{i,2},
\end{equation}
which gives the magnetic charges (cfr.\! eqn.\eqref{magch1})
\begin{equation}\label{didi2}
m_j(X)=\dim X_{j,1}-\dim X_{j,2}\qquad j=1,2,\dots, r,
\end{equation}
which are consistent with Dirac integrality \cite{cattoy,half}.
\medskip

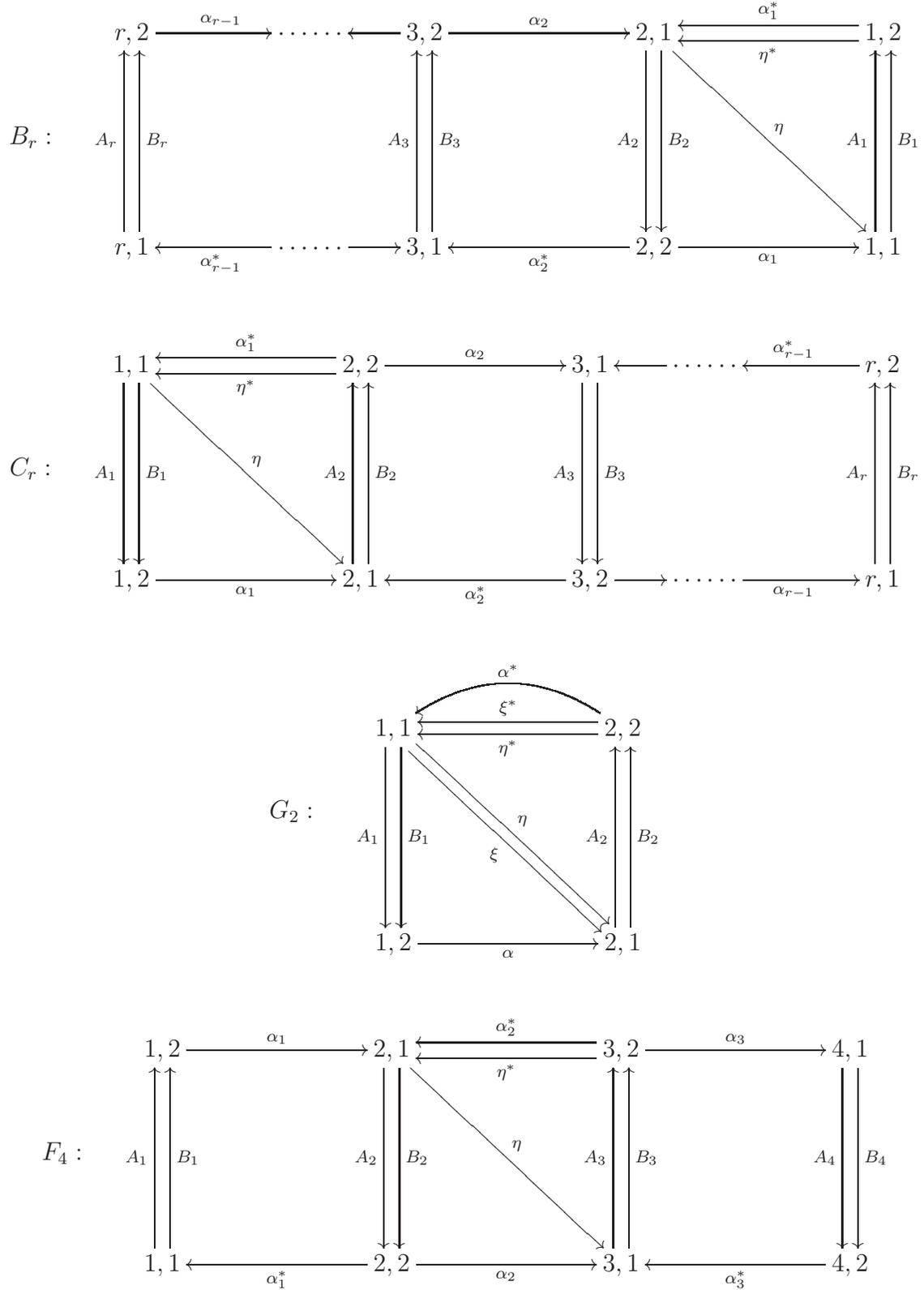
\begin{figure}
\begin{gather*}
B_r:\quad
\begin{gathered}
\xymatrix{r,2\ar[rr]^{\alpha_{r-1}}&&\cdots\cdots& 3,2\ar[rrr]^{\alpha_2}\ar[l]&&&2,1\ar[dddrrr]^{\eta}\ar@<0.7ex>[ddd]^{B_2}\ar@<-0.7ex>[ddd]_{A_2} 
&&& 1,2\ar@<0.7ex>[lll]^{\eta^*}\ar@<-0.7ex>[lll]_{\alpha_1^*} \\
\\
\\
r,1\ar@<0.7ex>[uuu]^{A_{r}}\ar@<-0.7ex>[uuu]_{B_{r}}&&\ar[ll]^{\alpha_{r-1}^*}\cdots\cdots\ar[r]&3,1\ar@<0.7ex>[uuu]^{A_3}\ar@<-0.7ex>[uuu]_{B_3}  &&& 2,2\ar[lll]^{\alpha_2^*}\ar[rrr]_{\alpha_1} &&& 1,1\ar@<0.7ex>[uuu]^{A_1}\ar@<-0.7ex>[uuu]_{B_1}}
\end{gathered}\\
\phantom{BBBBB}\\
C_r:\quad\begin{gathered}
\xymatrix{1,1\ar[dddrrr]^{\eta}\ar@<0.7ex>[ddd]^{B_1}\ar@<-0.7ex>[ddd]_{A_1} 
&&& 2,2\ar@<0.7ex>[lll]^{\eta^*}\ar@<-0.7ex>[lll]_{\alpha_1^*}\ar[rrr]^{\alpha_2}&&& 3,1\ar@<0.7ex>[ddd]^{B_3}\ar@<-0.7ex>[ddd]_{A_3} &\ar[l] \cdots\cdots && r,2\ar[ll]_{\alpha^*_{r-1}} 
  \\
\\
\\
1,2\ar[rrr]_{\alpha_1} &&& 2,1\ar@<0.7ex>[uuu]^{A_2}\ar@<-0.7ex>[uuu]_{B_2} &&&3,2\ar[r]\ar[lll]^{\alpha_2^*} &\cdots\cdots\ar[rr]_{\alpha_{r-1}}&&r,1\ar@<0.7ex>[uuu]^{A_r}\ar@<-0.7ex>[uuu]_{B_r} }
\end{gathered}
\\
\phantom{BBBBB}\\
G_2:\quad\begin{gathered}
\xymatrix{1,1
\ar@<0.5ex>[dddrrr]^{\eta}\ar@<-0.5ex>[dddrrr]_{\xi}\ar@<0.7ex>[ddd]^{B_1}\ar@<-0.7ex>[ddd]_{A_1} 
&&& 2,2\ar@<0.4ex>[lll]^{\eta^*}\ar@<-0.7ex>[lll]_{\xi^*}\ar@/_1.8pc/[lll]_{\alpha^*} \\
\\
\\
1,2\ar[rrr]_{\alpha} &&& 2,1\ar@<0.7ex>[uuu]^{A_2}\ar@<-0.7ex>[uuu]_{B_2}}
\end{gathered}
\\
\phantom{BBBBB}\\
F_4:\quad\begin{gathered}
\xymatrix{1,2\ar[rrr]^{\alpha_1}&&&2,1\ar[dddrrr]^{\eta}\ar@<0.7ex>[ddd]^{B_2}\ar@<-0.7ex>[ddd]_{A_2} 
&&& 3,2\ar@<0.7ex>[lll]^{\eta^*}\ar@<-0.7ex>[lll]_{\alpha_2^*}\ar[rrr]^{\alpha_3}&&& 4,1\ar@<0.7ex>[ddd]^{B_4}\ar@<-0.7ex>[ddd]_{A_4} 
  \\
\\
\\
1,1\ar@<0.7ex>[uuu]^{A_1}\ar@<-0.7ex>[uuu]_{B_1}&&&\ar[lll]^{\alpha_1^*}2,2\ar[rrr]_{\alpha_2} &&& 3,1\ar@<0.7ex>[uuu]^{A_3}\ar@<-0.7ex>[uuu]_{B_3} &&&4,2\ar[lll]^{\alpha_3^*} }
\end{gathered}
\end{gather*}
\caption{The quivers of SYM with non--simply laced gauge group. The corresponding superpotentials are given in section \ref{SEc:wwww}.}
\label{nonquivers}
\end{figure}

The quivers for SYM with $\mathfrak{g}=BCFG$ are then as in figure \ref{nonquivers}.

\medskip

 It remains to determine the superpotential $\cw(\mathfrak{g})$ such that the light subcategory $\mathscr{L}(\mathfrak{g})\subset\mathsf{rep}(Q(\mathfrak{g}),\cw(\mathfrak{g}))$ has the required physical properties \cite{cattoy} and, moreover, it enjoys the monodronic properties discussed in the previous section, see eqn.\eqref{uuuwer}. Before doing that, we review some facts about such light subcategories in a form convenient for our present purposes.

\section{Review of the light category $\mathscr{L}\subset \mathsf{rep}(Q,\cw)$}\label{reviewl}

Suppose we have a quiver $Q$ which corresponds to a $\cn=2$ theory that, in some corner of its parameter space, is a weakly coupled gauge theory with gauge group $G$. If we know the representations of $Q$ corresponding to the simple--roots $\alpha_i$, we may define the magnetic charges $m_i(X)\in \Z^r$ of all representations of $Q$ as in eqns.\eqref{didi1}\eqref{didi2}.
In the weak coupling limit $g_\mathrm{YM}\rightarrow 0$, the central charge of the representation $X$ behaves as \cite{cattoy}
\begin{equation}
Z(X)= -\frac{1}{g^2_\mathrm{YM}}\sum_i C_i\, m_i(X)+O(1),\qquad C_i>0,
\end{equation} 
so that all states which have bounded mass as  $g_\mathrm{YM}\rightarrow 0$ must have\footnote{\ To avoid technical complications, one may take the real numbers $C_i$ to be linearly independent over $\mathbb{Q}$.} $m_i(X)=0$ for all $i$. A representation $X$ with $m_i(X)=0$ will be very unstable if it has a subrepresentation $Y$ with $m_i(Y)>0$ for some $i$. This (and some mathematical theory \cite{cattoy}) motivates the definition of the light subcategory $\mathscr{L}$ of $\mathsf{rep}(Q,\cw)$ as the full subcategory over the objects $X$ satisfying the following two conditions:
\begin{align}
&(1)\qquad m_i(X)=0\ \ \text{for all }i=1,2,\dots,r,\\
&(2)\qquad \text{for all subobjects }Y\ \text{of }X,\ m(Y)_i\leq 0\ \text{for all }i.
\end{align} 
One shows that $\mathscr{L}$ is an exact Abelian subcategory which is closed under directs sum/summands and extensions.

The basic property of $\mathscr{L}$ is that it contains all representations corresponding to the stable BPS particles which are light in the region $g_\mathrm{YM}\sim 0$. If, in this regime, the theory has a weakly coupled Lagrangian description, the stable states of $\mathscr{L}$ (at weak coupling) precisely correspond to the perturbative spectrum of the model. In this case, on physical grounds, the light stable states should consist in one vector--multiplet in the adjoint of $\mathfrak{g}$ plus finitely many hypers in definite $\mathfrak{g}$--representations.  

Such physical considerations imply other Representation Theoretical properties of the light category $\mathscr{L}$ which may be rather surprising from the mathematical side. For instance, a perturbative state certainly cannot, at weak coupling, become non--perturbative by tuning the adjoint Higgs background to trigger the breaking $G\rightarrow SU(2)\times U(1)^{r-1}$. For a quiver as in figure \ref{nonquivers}, this means that\footnote{\ For the corresponding statement in more general situations, see \cite{half}.}
\begin{equation}\label{pain}
X\in\mathscr{L}\quad\Longrightarrow\quad X\big|_{\mathbf{Kr}_i}\in \mathscr{L}(SU(2))\quad \text{for all }i=1,2,\dots,r,
\end{equation}
where $\mathscr{L}(SU(2))$ denotes the light category for pure $SU(2)$ SYM, equal to the category of regular representations of the Kronecker quiver $\upuparrows$ \cite{cattoy,RI,ASS,CB}. With some pain the property \eqref{pain} may be proven directly from representation theory, see \cite{cattoy,half} and appendix \ref{app:lemma}.
\smallskip

As explained in refs.\!\cite{cattoy,half}, the property \eqref{pain} implies a major simplification; 
indeed, it allows to identify the sink $i,2$ and source $i,1$ nodes of each $\mathbf{Kr}_i$ subquiver by setting equal to 1 one of the arrows, say $B_i$, in each vertical Kronecker subquiver (cfr.\! figure \ref{nonquivers}). In this way we get a smaller quiver $Q^\prime$ having $r$ less nodes than the original one. Under this reduction, the superpotential $\cw$ induces an effective superpotential $\cw^\prime$ for $Q^\prime$ \cite{half} and one has simply  
$$\mathscr{L}=\mathsf{rep}(Q^\prime,\cw^\prime).$$
Notice that in a theory with a Lagrangian description all states in $\mathscr{L}$ are mutually local so that
\begin{equation}
\langle\cdot,\cdot\rangle_\text{Dirac}\big|_{\mathscr{L}}=0,
\end{equation}
and the net number of arrows between any two nodes of $Q^\prime$ must be zero
\begin{equation}
\#\{\text{arrows }\ i \rightarrow j \ \text{in }Q^\prime\}-\#\{\text{arrows }\ i \leftarrow j \ \text{in }Q^\prime\}=0.
\end{equation}
In particular, $Q^\prime$, if connected, cannot be 2--acyclic as the typical quiver for a 4d $\cn=2$ theory. This is related to the fact that $\mathscr{L}$ is not the non--perturbative Abelian category representing all BPS states of the theory, but rather it represent only a subsector (the perturbative one) which is not a full QFT. The complete spectrum also contain dyons with masses of order $O(1/g^2_\mathrm{YM})$. Lifting the perturbative category $\mathscr{L}$, which captures the classical physics, to its non--perturbative completion $\mathsf{rep}(Q,\cw)$ is a sort of quantization procedure which we shall refer to as `categorical quantization'. 
\smallskip

There is another property of the light category which is crucial for our purposes and was already mentioned in section \ref{sec:geometric}. The pure $SU(2)$ light category, $\mathscr{L}(SU(2))$, has the following structure \cite{RI,ASS,CB}: there is a family of orthogonal subcategories $\mathscr{L}(SU(2))_\lambda\subset \mathscr{L}(SU(2))$, labelled by a point $\lambda\in\mathbb{P}^1$ such that each object $X\in\mathscr{L}(SU(2))$ is a (finite) direct sum $\oplus_\lambda X_\lambda$ with $X_\lambda\in\mathscr{L}(SU(2))_\lambda$. In facts, $\mathscr{L}(SU(2))_\lambda$ is the closure with respect to direct sums and extensions of the brick representation
\begin{equation}
\begin{gathered}
\xymatrix{\C\ar@<0.7ex>[rr]^{\lambda}\ar@<-0.7ex>[rr]_{1} && \C}
\end{gathered}
\end{equation}
of dimension $\delta$ equal to the minimal imaginary root of $\widehat{A}_1$, which is physically identified with the charge vector of the $SU(2)$ $W$--boson. In symbols these facts are expressed as \cite{RI}
\begin{equation}\label{ringel}
\mathscr{L}(SU(2))=\bigvee_{\lambda\in\mathbb{P}^1}\mathscr{L}(SU(2))_\lambda.
\end{equation} 
Physically, this equations just states that the $W$--boson is a vector--multiplet and that there are neither light hypermultiplets nor
light higher--spin supermultiplets in the  spectrum of pure $SU(2)$ SYM \cite{cattoy}.

For general $\cn=2$ gauge theories eqn.\eqref{ringel} generalizes as follows \cite{cattoy}: the light category has the form
\begin{equation}
\mathscr{L}=\bigvee_{\lambda\in N}\mathscr{L}_\lambda,
\end{equation}
where $N$ is a one--dimensional variety whose irreducible components $N_a$ are in one--to--one correspondence with the simple factors in the gauge group $G$. Moreover, in class--$S$ theories, $N$ may be identified with the skeleton of the particular degeneration of the Gaiotto curve which is relevant for the $S$--duality frame specified by the given choice of magnetic charges $m_i(\cdot)$. Matter lives in the categories $\mathscr{L}_{\lambda_s}$ where $\{\lambda_s\}$ is a finite set of points in $N$, and the matter in $\mathscr{L}_{\lambda_s}$ may be charged only with respect to the gauge factor groups $G_a$ whose irreducible component
$N_a\ni \lambda_s$. Again, these statements just correspond to the fact that the gauge particles are vectors while the matter has only spin $0,1/2$ states. In the next section we shall show directly (and rigorously) that all these general statements hold true for the models (and categories) of interest in this paper. 
\smallskip

The situations explicitly analyzed in \cite{cattoy} correspond to the 
\emph{split}, \textit{i.e.}\! $ADE$ case. In that case, assuming the gauge group $G$ to be simple, eqn.\eqref{pain} has a nice refinement: for all $\lambda\in\mathbb{P}^1$,
\begin{equation}\label{pain2}
X\in\mathscr{L}_\lambda\quad\Longrightarrow\quad X\big|_{\mathbf{Kr}_i}\in \mathscr{L}(SU(2))_\lambda\quad \text{for all }i=1,2,\dots,r.
\end{equation}

This last equation will not longer be true in the \emph{non--split} case, and its failure is an alternative way of characterizing the non--split phenomenon.
\smallskip

It will be shown in the next two sections that the generalization of eqn.\eqref{pain2} to the case of a gauge group $G$ which is simple and \emph{non}--simply laced is as follows. Let $d_i\in\mathbb{N}$ be the integers defined in eqn.\eqref{defdi}. Then eqn.\eqref{pain2} is replaced by
\begin{equation}\label{pain3}
X\in\mathscr{L}_\lambda\quad\Longrightarrow\quad X\big|_{\mathbf{Kr}_i}\in \mathscr{L}(SU(2))_{\lambda^{1/d_i}}\quad \text{for all }i=1,2,\dots,r.
\end{equation}
The fact that in the \textsc{rhs} we have to choose a square root (or a cubic root in the $G_2$ case) of $\lambda\in \C^\times$ opens the possibility that going along a non--contractible loop in $\C^\times$ we end up with a different choice for the root. This will produce a monodromy having the form predicted in eqn.\eqref{uuuwer}. More detailed statements below.

\section{The light category of non--simply laced $\cn=2$ SYM}

As in ref.\!\cite{half}, our strategy for fixing the superpotential $\cw(\mathfrak{g})$ is to require that the corresponding category $\mathsf{rep}(Q(\mathfrak{g}),\cw(\mathfrak{g}))$ contains a light category $\mathscr{L}(\mathfrak{g})$ with the right physical properties as described in \S.\,\ref{reviewl}, and, in particular, that the light BPS states at weak coupling consist of just one vector--multiplet in the adjoint of $\mathfrak{g}$. Contrary to ref.\!\cite{half}, however, we shall not put forward an ansatz for $\cw(\mathfrak{g})$ and then check the properties of $\mathscr{L}(\mathfrak{g})$, but rather we shall directly construct the correct light category $\mathscr{L}(\mathfrak{g})$ from the monodromic properties it must enjoy (section \ref{sec:geometric}), and then build the full non--perturbative  category 
$\mathsf{rep}(Q(\mathfrak{g}),\cw(\mathfrak{g}))$  out of its light subcategory $\mathscr{L}(\mathfrak{g})$ by the process 
we have dubbed 
 `categorical quantization'.
 \medskip
 
 We start by constructing the non--simply laced monodromic counterparts to the usual $ADE$ preprojective algebras \cite{pre1,pre2,CBlemma,GLS} which played such a prominent role in refs.\!\cite{cattoy,half}.

\subsection{Monodromic light categories}

We start by considering certain monodromic module categories which play for the non--simply laced case a role somehow similar to that played, in the simply laced case \cite{cattoy}, by the modules over the preprojective algebra $\cp(G)$ (where the graph $G$ is the Dynkin diagram of the gauge group).

\subsubsection{Preprojective algebras for non--simply laced (\textit{i.e.\!} valued) graphs}\label{sec:fghat}

For our present purposes, a \emph{valued} graph $\mathfrak{g}$ is a \emph{simply laced} graph together with the assignment of a pair $(n_\alpha,m_\alpha)$ of positive integers to each link $\alpha$  of $\mathfrak{g}$; at least one of the two numbers $(n_\alpha,m_\alpha)$ must be 1, and if they are both 1 the link is said to have trivial valuation and is denoted by a plain edge. The Dynkin graph of a finite--dimensional (simple) Lie algebra $\mathfrak{g}$ may be seen as a valued graph with valuation function
\begin{equation}
\big(n(\alpha),m(\alpha)\big)\equiv \Big(|C_{s(\alpha),t(\alpha)}|,|C_{t(\alpha),s(\alpha)}|\Big).
\end{equation}
$\mathfrak{g}$ is simply laced precisely iff all links have trivial valuation.\smallskip

Given a valued graph $\mathfrak{g}$, we define its \textit{$N=2$ double} 
$\widehat{\mathfrak{g}}$ to be the quiver obtained by replacing 
 each edge $\alpha$ of $\mathfrak{g}$ with a pair of opposite oriented arrows\footnote{\ We shall refer to the arrows $\alpha$ as the \emph{direct} arrows, while the arrows $\alpha^*$ will be called \emph{inverse} arrows.} 
$$\xymatrix{\ar@<0.6ex>[rr]^{\alpha_l}&&\ar@<0.6ex>[ll]^{\alpha_l^*}}$$ 

In addition, the quiver $\widehat{\mathfrak{g}}$ also has an `adjoint' loop $A_i$ at each node $i$ of $\mathfrak{g}$. For future convenience, between nodes $i$, $j$ we also draw
$(\max\{|C_{ij}|,|C_{ji}|\}-1)$ \emph{dashed} pairs of opposite arrows. We stress that the dashed arrows are NOT part of the quiver $\widehat{\mathfrak{g}}$, and most of the times we shall even \emph{omit} to draw them; they will be needed only in \S.\,\ref{auxaux}. \smallskip

For instance, for $\mathfrak{g}=F_4$ we get
\begin{equation}
\begin{gathered}
\widehat{{F_4}}:\quad \xymatrix@R=2.0pc@C=3.0pc{
&\ar@(ul,dl)[]_{A_1} 1\ar@<0.6ex>[rr]^{\alpha_1}&& 2\ar@<0.4ex>[rr]^{\alpha_2} \ar@<0.7ex>[ll]^{\alpha_1^*}\ar@(dl,dr)[]_{A_2}  \ar@{..>}@<-1ex>@/_1.5pc/[rr]_{\eta}  &&\ar@{..>}@<-1ex>@/_1.5pc/[ll]_{\eta^*}\ar@<0.4ex>[ll]^{\alpha_2^*} \ar@(ur,ul)[]_{A_3} 3  \ar@<0.6ex>[rr]^{\alpha_3} && \ar@<0.6ex>[ll]^{\alpha_3^*} 4 \ar@(ru,rd)[]^{A_4} 
}
\end{gathered}
\end{equation}

We equip $\widehat{\mathfrak{g}}$ with the superpotential\footnote{\ Here and in the following we adopt the more concise mathematical notation for the superpotential \cite{derksen1}. Under the identification of the arrows of the quiver with the 1d `bifundamental' Higgs fields, the physical superpotential is $$W=\mathrm{tr}\,\cw.$$
Therefore, two words of $\C Q$, corresponding to an oriented cycle in $Q$, are considered equivalent as contributions to $\cw$ if they differ by a cyclic permutation of the arrows.}
\begin{equation}\label{msuper}
\mathscr{W}=\sum_\alpha \Big( \alpha\, A_{s(\alpha)}^{n(\alpha)}\,\alpha^*-\alpha^*\, A_{t(\alpha)}^{m(\alpha)}\,\alpha\Big),
\end{equation}
where the sum is over the direct arrows only.\smallskip

We define the algebra $\cp(\mathfrak{g})$ to be the Jacobian algebra
\begin{equation}
\cp(\mathfrak{g}):=\C\, \widehat{\mathfrak{g}}/(\partial\mathscr{W}).
\end{equation}

In case $\mathfrak{g}$ is simply laced (\textit{i.e.}\! trivially valued), $\cp(\mathfrak{g})$ is essentially the preprojective algebra of the graph $\mathscr{g}$ \cite{pre1,pre2,CBlemma,GLS}. We shall loosely refer to $\cp(\mathfrak{g})$ as \textit{the preprojective algebra} of $\mathfrak{g}$ also in the non--simply laced case.

\subsubsection{The structure of $\mathsf{mod}\,\cp(\mathfrak{g})$}\label{structurecat}

From now on the valued graph $\mathfrak{g}$ is the Dynkin diagram of a finite--dimensional simple Lie algebra. 

We wish to prove the following two crucial facts:\medskip

\textit{(1) The Abelian category $\mathsf{mod}\,\cp(\mathfrak{g})$ has a decomposition into (exact, full, extension--closed) Abelian subcategories of the following form
\begin{equation}
\mathsf{mod}\,\cp(\mathfrak{g})\equiv \mathscr{M}(\frak{g})=\bigvee_{\lambda\in \mathbb{P}^1} \mathscr{M}(\frak{g})_\lambda.
\end{equation}}

\textit{(2) The bricks of $\cp(\mathfrak{g})$ belong to $\mathbb{P}^1$ families $X_\lambda\in\mathscr{M}(\frak{g})_\lambda$ ($\lambda\in \mathbb{P}^1$) and their dimension vectors are positive roots of $\mathfrak{g}$.}
\medskip

These results are known to hold in the simply--laced cases and, in that situation, they imply that in each weakly coupled chamber the light BPS spectrum consists precisely of vector--multiplets making one copy of the adjoint of $\mathfrak{g}$ \cite{cattoy}. However, in the non--simply laced case, the corresponding representations will also have non--trivial monodromy, in agreement with the geometric discussion of section \ref{sec:geometric}.
\medskip

Again we write $d_i$ ($i=1,2,\dots, r$) for the integers defined in eqn.\eqref{defdi}.  
The valuation of the edge $\xymatrix{i \ar@{-}[r]^\alpha&j}$ of $\mathfrak{g}$ is
\begin{equation}
\big(n(\alpha),\: m(\alpha)\big)= \left(\frac{d_i}{(d_i,d_j)},\: \frac{d_j}{(d_i,d_j)}\right).
\end{equation}

As a consequence of the relations $\partial\mathscr{W}=0$, for all $X\in\mathsf{mod}\,\cp(\mathfrak{g})$ the set of linear maps
\begin{equation}
X_i\longmapsto A_i^{d_i}X_i\quad \text{for all }i,
\end{equation} 
gives an element of $\text{End}\,X$. Suppose $X$ is \emph{indecomposable}. Then $\text{End}\,X$ is a \emph{local} ring \cite{ASS1},
and hence there is a complex number $\lambda\in\C$ and integers $f_i$ such that
\begin{equation}\label{krull}\big(A_i^{d_i}-\lambda\big)^{f_i}=0\quad \text{for all }i=1,2,\dots, r.
\end{equation}
We write $\mathscr{M}(\frak{g})_\lambda$ for the Krull--Schimdt full subcategory of $\mathsf{mod}\,\cp(\mathfrak{g})$ whose objects are the direct sums of indecomposables satisfying \eqref{krull} for the given $\lambda\in\C$. It is manifestly an Abelian category. This shows (1).\medskip

In the particular case that $X$ is actually a \text{brick} we have the stronger statement
\begin{equation}
X\ \text{is a brick}\quad \Longrightarrow\quad A_i^{d_i}=\lambda\ \ \text{for all }i=1,2,\dots, r.
\end{equation}
Let us consider the case of $\lambda$ generic (\textit{i.e.}\! $\lambda\neq 0$). Then $A_i$ is semisimple with eigenvalues
$e^{2\pi i \ell_i/d_i}\,\lambda^{1/d_i}$, $\ell_i=1,2,\dots, d_i$.
Let $P_{i,\ell_i}$ be the projector on the corresponding eigenspace, so that
\begin{equation}
X_i=P_{i,1}X_i\oplus \cdots \oplus P_{i,d_i}X_i.
\end{equation}
Thus we split each vector space $X_i$ in as many vector spaces  as the number of nodes of the parent graph $\mathfrak{g}_\mathrm{parent}$ folded into the $i$--th node. By projecting the arrows in these eigenspaces, we get a representation $X^\flat$ of the double quiver $\overline{\mathfrak{g}_\text{parent}}$ which satisfies the usual $\cp(\mathfrak{g}_\text{parent})$ relations. It is elementary to check that $X$ is a brick iff the $\cp(\mathfrak{g}_\text{parent})$ module $X^\flat$ is a brick. Then the statement reduces to the simply--laced case, which is known to hold (\!\cite{cattoy} and reference threin). Thus (2) is true.

\medskip
A few explicit examples will make (1)(2) more concrete.

\subsubsection{Example: $B_2$}\label{exb2}
In this case the quiver and superpotential are
\begin{align}\label{quqieeie}
\begin{gathered}
\xymatrix@R=2.0pc@C=3.0pc{
&\ar@(ul,dl)[]_{A_1} 1  \ar@<-0.8ex>[rrr]_{\alpha} &&&\ar@<-0.8ex>[lll]_{\alpha^*}  \ar@(ur,dr)[]^{A_2}2 
}
\end{gathered}
\end{align}
\begin{equation}
\mathscr{W}=\alpha\, A_1\alpha^* -\alpha\, \alpha^* A_2^2.
\end{equation}
The relations are
\begin{align}
&\partial_\alpha \mathscr{W}\equiv A_1\alpha^*-\alpha^* A_2^2=0\label{rel1X}\\
& \partial_{\alpha^*}\mathscr{W}\equiv \alpha\, A_1-A_2^2\,\alpha=0\label{rel2X}\\
&\partial_{A_1}\mathscr{W}\equiv \alpha^*\alpha=0\label{rel3X}\\
&\partial_{A_2}\mathscr{W}\equiv - \alpha\,\alpha^* A_2-A_2\,\alpha\,\alpha^*=0.\label{rel4X}
\end{align}
from which we see that the map
\begin{equation}
\ell\colon\quad (X_1,X_2)\mapsto (A_1X_1,A_2^2X_2),
\end{equation}
is an element of $\text{End}\,X$ and hence a complex number $\lambda\in \mathbb{P}^1$ if $X$ is a brick.

For $\lambda$ generic (that is, $\lambda\neq 0,\infty$)
$A_2$ may be written in the form
\begin{equation}
A_2= \sqrt{\lambda}\left(\begin{array}{c|c} \boldsymbol{1}_{n\times n}& \boldsymbol{0}_{n\times m}\\\hline
\boldsymbol{0}_{m\times n} & -\boldsymbol{1}_{m\times m}\end{array}\right).
\end{equation}
 Let $P_\pm$ the projectors to the two eigenspaces of $A_2$. Clearly a brick $X$ with $\lambda\neq 0,\infty$ induces a representation of the \emph{unfolded} quiver
\begin{align}
&\begin{gathered}
\xymatrix@R=2.0pc@C=3.0pc{
&P_+X_2  \ar@/_1.6pc/[rrr]_{\alpha^* P_+} &&&\ar@/_1.6pc/[lll]_{P_+\alpha} X_1  \ar@/_1.6pc/[rrr]_{P_-\alpha } &&&\ar@/_1.6pc/[lll]_{-\alpha^* P_-} P_-X_2  
}
\end{gathered}
\end{align}
which satisfies the relations
\begin{align}
& P_\pm\alpha \cdot\alpha^* P_\pm=0 && \text{\begin{footnotesize}(by eqn.\eqref{rel4X})\end{footnotesize}}\\
& \alpha^* P_+\cdot P_+\alpha-(-\alpha^* P_-)\cdot P_-\alpha = \alpha^*\,\alpha=0,&& \text{\begin{footnotesize}(by eqn.\eqref{rel3X})\end{footnotesize}}
\end{align} 
which are precisely those of $\cp(A_3)$.
Hence the bricks of $X$ are mapped into representations of $\cp(A_3)$ which have dimension vectors equal to the positive roots of $A_3$. By folding back the quiver, we get (essentially by definition) the positive roots of $B_2$. 

\subsubsection{Example: $G_2$} 
The quiver is the same as in eqn.\eqref{quqieeie} with a different superpotential
\begin{equation}
\mathscr{W}=\alpha\, A_1\,\alpha^* - \alpha\, \alpha^*\, A_2^3.
\end{equation}
The relations are
\begin{align}
&\partial_\alpha \mathscr{W}\equiv A_1\,\alpha^*-\alpha^*\, A_2^3=0\label{rel13X}\\
& \partial_{\alpha^*}\mathscr{W}\equiv \alpha\, A_1- A_2^3\,\alpha=0\label{rel23X}\\
&\partial_{A_1}\mathscr{W}\equiv \alpha^8\, \psi=0\label{rel33X}\\
&\partial_{A_2}\mathscr{W}\equiv -(A_2^2\,\alpha\,\alpha^*+A_2\,\alpha\,\alpha^*\, A_2+\alpha\,\alpha^*\, A_2^2)=0.\label{rel43X}
\end{align}
The map
\begin{equation}
(X_1, X_2)\mapsto (A_1X_1,A_2^3X_2)
\end{equation}
is an element of $\mathrm{End}\,X$, hence a complex number $\lambda$.
For generic $\lambda$, $A_2$ has eigenvectors $\lambda\, e^{2\pi i k/3}$, with $k=0,1,2$. Let $P_0,P_1,P_2$ be the projectors on the corresponding eigenspaces. Then we get a representation of the quiver
\begin{align}
&\begin{gathered}
\xymatrix@R=2.0pc@C=3.0pc{
&P_0X_2  \ar@/_2pc/[rrr]_{\alpha^*\, P_0} &&&\ar@/_2pc/[lll]_{P_0\,\alpha} X_1   \ar@/_2pc/[ddd]_{P_1\,\alpha }\ar@/_2pc/[rrr]_{P_2\,\alpha } &&&\ar@/_2pc/[lll]_{\alpha^*\, P_2} P_2X_2  \\
\\
\\
&&&& P_1X_2\ar@/_2pc/[uuu]_{\alpha^*\,P_1 }
}
\end{gathered}
\end{align}
which satisfies the relations of $\cp(D_4)$ (cfr.\! eqns.\eqref{rel43X}\eqref{rel33X})
\begin{align}
&P_k\alpha\cdot \alpha^* P_k=0 \qquad k=0,1,2
&\sum_k (\alpha^* P_k)\cdot (P_k\alpha)=0.
\end{align}

\subsubsection{Example: $F_4$}

The quiver and $\mathscr{W}$ in this case are
\begin{gather}\label{quqieeiexxx}
\begin{gathered}
\xymatrix@R=2.0pc@C=3.0pc{
&\ar@(ul,dl)[]_{A_1} 1  \ar@<-0.8ex>[rr]_{\alpha_1} &&\ar@<-0.8ex>[ll]_{\alpha^*_1}  \ar@(ur,ul)[]_{A_2}2 \ar@<-0.8ex>[rr]_{\alpha_2} &&\ar@<-0.8ex>[ll]_{\alpha^*_2}  \ar@(ur,ul)[]_{A_3}3  \ar@<-0.8ex>[rr]_{\alpha_3} &&\ar@<-0.8ex>[ll]_{\alpha^*_3}  \ar@(rd,ru)[]_{A_4}4
}
\end{gathered}\\
\mathscr{W}=A_1\,\alpha^*_1\,\alpha_1-A_2\,\alpha_1\,\alpha^*_1+A_2\,\alpha^*_2\,\alpha_2-
A_3^2\,\alpha_2\,\alpha^*_2+A_3\,\alpha^*_3\,\alpha_3-A_4\,\alpha_3\,\alpha^*_3.
\end{gather}
From the relations $\partial_{\alpha^*_i}\mathscr{W}=\partial_{\alpha_i}\mathscr{W}=0$ it follows that the map
\begin{equation}
(X_1,X_2,X_3,X_4)\longmapsto (A_1X_1,A_2X_2,A_3^2X_3,A_4^2X_4),
\end{equation}
is an element of $\mathrm{End}\,X$ and hence a number $\lambda\in \C$ if $X$ is a brick. Then $A_3$ and $A_4$ have eigenvalues $\pm \sqrt{\lambda}$, and (for $\lambda\neq0$) let $P_\pm$ and $Q_\pm$ be the corresponding projectors. We get the representation
\begin{equation}\label{e6f4}
\begin{gathered}
\xymatrix{&&&&P_+X_3 \ar@<0.6ex>[rrr]^{Q_+\alpha_3 P_+/(2\sqrt{\lambda})}\ar@<0.6ex>[dll]^{\alpha^*_2P_+} &&& Q_+X_4 \ar@<0.6ex>[lll]^{P_+\alpha^*_3 Q_+}\\
X_1\ar@<0.6ex>[rr]^{\alpha_1}&& X_2\ar@<0.6ex>[ll]^{\alpha^*_1}\ar@<0.6ex>[rru]^{P_+\alpha_2}\ar@<0.6ex>[rrd]^{P_-\alpha_2} \\
&&&&P_-X_3 \ar@<0.6ex>[ull]^{\alpha^*_2P_-}\ar@<0.6ex>[rrr]^{-Q_-\alpha_3 P_-/(2\sqrt{\lambda})} &&&Q_-X_4\ar@<0.6ex>[lll]^{P_-\alpha^*_3 Q_-}}\end{gathered}
\end{equation}
Notice that the relations imply
\begin{equation}
P_\pm \alpha^*_3 Q_\mp=Q_\pm \alpha_3P_\mp=0,
\end{equation}
so the arrows connect the various subspaces precisely as in the above quiver \eqref{e6f4}. The relations of the projective algebra $\cp(E_6)$ are then satisfied; for instance
\begin{equation}
P_\pm \alpha_2\cdot\alpha^*_2P_\pm - 
P_\pm \alpha^*_3 Q_\pm\cdot \left(\pm \frac{1}{2\sqrt{\lambda}}Q_\pm \alpha_3P_\pm \right)=\mp \frac{1}{2\sqrt{\lambda}}P_\pm\; \partial_{A_3}\mathscr{W}\, P_\pm=0.
\end{equation}
Again our claim is verified. 

\subsubsection{Monodromy of $\mathscr{M}(\mathfrak{g})$}

From the above examples, it is clear that there is an auto--equivalence
\begin{equation}
\mathscr{G}\colon \mathscr{M}(\mathfrak{g})_\lambda\rightarrow \mathscr{M}(\mathfrak{g})_\lambda 
\end{equation}
which acts by permuting the roots of the algebraic equation
\begin{align*}&x^2=\lambda &&\text{for } \mathfrak{g}\neq G_2\\ &x^3=\lambda && \text{for } \mathfrak{g}=G_2.\end{align*}

For instance, $\mathscr{G}$ will act on a brick in $\mathscr{M}(G_2)_\lambda$ with dimension $(1,1)$ as
\begin{align}
\begin{gathered}
\xymatrix@R=2.0pc@C=3.0pc{
&\ar@(ul,dl)[]_{\lambda} \C  \ar@<-0.8ex>[rr] &&\ar@<-0.8ex>[ll]  \ar@(ur,dr)[]^{\lambda^{1/3}}\C 
}
\end{gathered}\ \xrightarrow{\ \ \mathscr{G}\ \ }\hskip-0.5cm
\begin{gathered}
\xymatrix@R=2.0pc@C=3.0pc{
&\ar@(ul,dl)[]_{\lambda} \C  \ar@<-0.8ex>[rr] &&\ar@<-0.8ex>[ll]  \ar@(ur,dr)[]^{e^{2\pi i/3}\,\lambda^{1/3}}\C 
}
\end{gathered}
\end{align}
which corresponds in terms of the covering $\cp(D_4)$ representations to
\begin{equation}
\begin{gathered}
\xymatrix{& \C \ar@<0.4ex>[dl]\\
\C \ar@<0.5ex>[ur]\ar@<0.4ex>[r]\ar@<0.4ex>[dr] & 0 \ar@<0.4ex>[l]\\
& 0\ar@<0.5ex>[ul]}
\end{gathered}\quad\xrightarrow{\quad\qquad}\quad 
\begin{gathered}
\xymatrix{& 0 \ar@<0.4ex>[dl]\\
\C \ar@<0.5ex>[ur]\ar@<0.4ex>[r]\ar@<0.4ex>[dr] & \C \ar@<0.4ex>[l]\\
& 0\ar@<0.5ex>[ul]}
\end{gathered}
\end{equation}

Then the family of Abelian categories $\mathscr{M}(\mathfrak{g})_\lambda$ manifestly enjoys the monodronic properties discussed in section \ref{sec:geometric}, namely
\begin{equation}
\mathscr{M}(\mathfrak{g})_{e^{2\pi i}\lambda}= \mathscr{G}\,\mathscr{M}(\mathfrak{g})_{\lambda}.
\end{equation}

Note that the action of $\mathscr{G}$ on the objects of $\mathscr{M}(\mathfrak{g})_\lambda$ (taken up to isomorphism) is \underline{not} free. For instance, in the $G_2$ case the simple representation of dimension $(1,0)$ is fixed by the functor
$\mathscr{G}$. More generally, under the lift
$$\mathscr{M}(\mathfrak{g})\rightarrow \mathsf{mod}\,\cp(\mathfrak{g}_\text{parent})$$ the action of $\mathscr{G}$ gets identified with the auto--equivalence $\varpi$ associated to the automorphisms of the Dynkin graph $\mathfrak{g}$. Then the action of $\mathscr{G}$ is not free precisely on the objects whose lift is $\varpi$--invariant.
\smallskip

Since the action of $\mathscr{G}$ is not free, the present construction differs from the notion of a Galois cover of a quiver as defined by Gabriel \cite{gab1,gab2}, which requires the group to act freely on the nodes of the quiver.

\subsection{`Auxiliary' fields and the final form of the light category
 $\mathscr{L}(\mathfrak{g})$}\label{auxaux}

In the definition of the $N=2$ double quiver $\widehat{\mathfrak{g}}$ in \S.\,\ref{sec:fghat} we added pairs $\xymatrix{\ar@<0.3ex>@{..>}[r] & \ar@<0.3ex>@{..>}[l]}$ of \emph{dashed} arrows which we dropped in all subsequent discussions. Now it is time to put them to some good use.\medskip

The point is that, for $\mathfrak{g}$ \emph{non}--simply laced, the superpotential $\mathscr{W}$ in eqn.\eqref{msuper} is not in a form suitable for categorical quantization. In facts, it is not linear in the loops $A_i$ --- that is, in the physical language, is not of first order in the adjoint Higgs fields $A_i$. This makes $1d$ \emph{extended} \textsc{susy} not manifest. There is a simple and well known remedy to this problem: We have to add to the 1d SQM massive `auxiliary' Higgs in such a way that integrating them away will produce the higher order term in the superpotential. The dashed arrows in $\widehat{\mathfrak{g}}$ correspond to such `auxiliary' Higgs fields. Then we shall describe the category $\mathscr{M}(\mathfrak{g})$, up to equivalence, in the form 
\begin{equation}
\mathscr{M}(\mathfrak{g})=\mathsf{rep}\big(Q(\mathfrak{g})^\prime,
\cw(\mathfrak{g})^\prime\big) \equiv \mathscr{L}(\mathfrak{g}),
\end{equation}
where the quiver $Q(\mathfrak{g})^\prime$ is just $\widehat{\mathfrak{g}}$ with the dashed arrows made solid and $\cw(\mathfrak{g})^\prime$ is such that it becomes $\mathscr{W}$ after integrating away the massive arrows, that is, in the mathematical language, $(\widehat{\mathfrak{g}}, \mathscr{W})$ is the \emph{reduced} pair of $(Q(\mathfrak{g})^\prime,
\cw(\mathfrak{g})^\prime)$ in the sense of DWZ \cite{derksen1}.
\smallskip

In the Dynkin graph of a non--simply laced finite--dimensional (simple) Lie algebra we have precisely \emph{one}
link with  a non--trivial valuation (which is (1,2), except for $G_2$ which has  a (1,3) edge). Therefore, in the sum in the \textsc{rhs} of eqn.\eqref{msuper} there is precisely one term of higher order in the loop arrows $A_i$. This is the only term we have to rewrite in terms of lower order ones. Therefore we may focus on the full subquiver of $Q(\mathfrak{g})^\prime$ over the two nodes connected by the link with the non--trivial valuation, since everything else will remain invariant.

We are reduced to consider two situations, corresponding to the two possible non--trivial link valuations. The relevant (sub)quiver in the two cases is, respectively,
\begin{equation}
\begin{gathered}
\xymatrix@R=2.0pc@C=3.0pc{
&\ar@(ul,dl)[]_{A_i} i \ar@<-1.4ex>@/_1.3pc/[rrr]_{\eta}  \ar@<-0.7ex>[rrr]_{\alpha} &&&\ar@<-0.7ex>[lll]_{\alpha^*}  \ar@(ur,dr)[]^{A_j} j\ar@<-1.4ex>@/_1.3pc/[lll]_{\eta^*} 
}
\end{gathered}
\end{equation}
\begin{equation}
\begin{gathered}
\xymatrix@R=2.0pc@C=3.0pc{
&\ar@(ul,dl)[]_{A_i} i \ar@<-2ex>@/_2.7pc/[rrr]_{\xi} \ar@<-1.4ex>@/_1.3pc/[rrr]_{\eta}  \ar@<-0.7ex>[rrr]_{\alpha} &&&\ar@<-0.7ex>[lll]_{\alpha^*}  \ar@(ur,dr)[]^{A_j} j\ar@<-1.4ex>@/_1.3pc/[lll]_{\eta^*}
\ar@<-2ex>@/_2.7pc/[lll]_{\xi^*} 
}
\end{gathered}
\end{equation}
while the relevant terms in $\mathscr{W}$ are (cfr.\! eqn.\eqref{msuper})
\begin{gather}\label{www1}
\mathscr{W}=A_i\,\alpha^*\,\alpha- A_j^2\,\alpha\,\alpha^*+\cdots\\
\mathscr{W}=A_i\,\alpha^*\,\alpha- A_j^3\,\alpha\,\alpha^*+\cdots
\label{www2}\end{gather}

Using the auxiliary arrows $\eta,\eta^*,\xi,\xi^*$, these terms are rewritten, respectively, as
\begin{gather}
\cw^\prime=A_i\,\alpha^*\,\alpha+A_j\alpha\,\eta^*+A_j\,\eta\,\alpha^*+\eta\,\eta^*+\cdots\\
\cw^\prime=A_i\,\alpha^*\,\alpha- A_j\,\alpha\,\xi^*-A_j\,\xi\,\eta^*-A_j\,\eta\,\alpha^*+\xi\,\xi^*+\eta\,\eta^*
+\cdots
\end{gather}
It is easy to check that integrating out the (massive) auxiliary arrows gives back \eqref{www1}\eqref{www2}. Using this rule, we construct $\cw(\mathfrak{g})^\prime$ for arbitrary (non--simply laced) $\mathfrak{g}$.
\bigskip

The conclusion of this section is that we have an equivalence
between the light category for $\mathfrak{g}$ $\cn=2$ SYM which (for the moment) we just \emph{define} to be
\begin{equation}\label{sumsu}\mathscr{L}(\mathfrak{g})\equiv\mathsf{rep}\big(Q(\mathfrak{g})^\prime, \cw(\mathfrak{g})^\prime\big)
\end{equation}
and the module category $\mathscr{M}(\mathfrak{g})$ we introduced previously. Then, using results (1)(2) of \S.\,\ref{structurecat}, we deduce the following crucial properties:
\begin{align}
&(1)\quad\text{($\mathbb{P}^1$--family)}
&& \mathscr{L}(\mathfrak{g})=\bigvee\nolimits_{\lambda\in \mathbb{P}^1}\mathscr{L}(\mathfrak{g})_\lambda\\
&(2)\quad\text{(non--split monodromy)}
&& \mathscr{L}(\mathfrak{g})_{e^{2\pi i}\lambda}=\mathscr{G}\,\mathscr{L}(\mathfrak{g})_\lambda\\
&(3)\quad\text{(bricks \& positive roots)}
&&X\ \text{is a brick }\Rightarrow\ \dim  X\in \Delta^+(\mathfrak{g}).\end{align}
In facts, (3) is slightly stronger than what we have actually proven, since we have shown it only for $\lambda\neq 0,\infty$. It is easy to see that the result extends to $\lambda=0$. For instance, in the example of \S.\,\ref{exb2}, at $\lambda=0$, $\alpha^*\,\alpha=0$ while the matrices
$\gamma_1=\alpha\,\alpha^*$ and $\gamma_2=A_2$ satisfy the Clifford algebra of the trivial quadratic form
\begin{equation}
\gamma_1^2=\gamma_2^2=\gamma_1\gamma_2+\gamma_2\gamma_1=0,
\end{equation} 
which, by the universality of the Clifford algebra, means that the representations of the quiver 
\begin{align}\label{quqieeie34}
\begin{gathered}
\xymatrix@R=2.0pc@C=3.0pc{
& 1  \ar@<-0.8ex>[rrr]_{\alpha} &&&\ar@<-0.8ex>[lll]_{\alpha^*}  \ar@(ur,dr)[]^{\gamma_2}2 
}
\end{gathered}
\end{align}
which are not direct sums of representations with either $\gamma_1$ or $\gamma_2$ vanishing, must contain a direct summand of the form
\begin{equation}
\alpha=\begin{pmatrix}1\\ 0\end{pmatrix}\otimes \mathbf{1}_2,\quad \alpha^*=\begin{pmatrix}0 & 1\end{pmatrix}\otimes \mathbf{1}_2,\quad A_2=\sigma_3\otimes \sigma^+,
\end{equation}
whose endomorphism ring contains the nilpotent $\mathbf{1}_2\otimes \sigma^+$. Hence, if  $X$ is  a brick,  either $\alpha\,\alpha^*$ or $A_2$ is zero. If $A_2=0$, we are reduced to a string algebra \cite{stringal} whose strings have dimensions in $\Delta^+(B_2)$. In the other case, $\alpha\,\alpha^*=\alpha^*\,\alpha=0$, and we again reduce to a string algebra whose strings have dimensions  in $\Delta^+(B_2)$ or equal to $(2,2)$, and in the last case $\mathrm{End}\,X$ necessarily contains nilpotent elements. In conclusion, all bricks in $\mathscr{L}(B_2)_0$ have dimension vectors in $\Delta^+(B_2)$. The opposite case $\lambda=\infty$ will be briefly commented in the next section.
\smallskip

From the physical point of view the fact that (3) extends to $\lambda=0,\infty$ is often obvious. Indeed, the existence of bricks at $\lambda=0,\infty$ with dimensions $\not\in\Delta^+(\mathfrak{g})$ would be interpreted as the coupling to SYM of matter which must be Lagrangian and carry no flavor charge, that is, to consist of \emph{half}--hypermultiplets. For most non--simply laced $\mathfrak{g}$  there is just no such matter systems.

\section{The quiver superpotential $\cw$ for $BCFG$ SYM}\label{SEc:wwww}

To complete the construction, we have to endow the quivers in figure 
\ref{nonquivers} with superpotentials $\cw(\mathfrak{g})$, then determine the corresponding light categories $\mathcal{L}(\mathfrak{g})$ --- which are well--defined since we know the magnetic charges, see section \ref{sec;dirac} --- and finally check that we obtain the same light categories as defined in eqn.\eqref{sumsu}. After that, we are guaranteed that our pairs
$(Q(\mathfrak{g}),\cw(\mathfrak{g}))$ predict the physically correct weak coupling spectrum, namely just one light BPS vector--multiplet in the adjoint of $\mathfrak{g}$, and hence the pair $(Q(\mathfrak{g}),\cw(\mathfrak{g}))$ is consistently identified with the quiver with superpotential for pure  $\cn=2$ $\mathfrak{g}$ SYM. On the other hand, these pairs precisely correspond to the geometrical split/non--split realization of $ADE$ \textit{vs.}\! $BCFG$ gauge symmetries, so if the $ADE$ pairs are correctly identified (and they are\,!!) so are the $BCFG$ ones.
\medskip

\smallskip

The quivers in figure \ref{nonquivers} by construction satisfy the conditions that the $X_{\alpha_i}$ have Kronecker supports; in facts they are precisely the representations with support in the vertical
Kroneckers $\upuparrows$ (resp.\! $\downdownarrows$) and dimension equal to the corresponding minimal imaginary root. Then the inverse process from the light effective superpotential $\cw^\prime$ to the full one $\cw$ consists in just reinserting the arrows $B_i$ (cfr.\! figure \ref{nonquivers} ) which were set to 1 in the reduction from the full non--perturbative module category to its light subcategory. There is a unique gauge--invariant way of doing this, so $\cw$ is uniquely recovered.
\medskip

With reference to the labeling of arrows in figure \ref{nonquivers}, the superpotentials are
\begin{align}
&\begin{aligned}\cw(B_r)&=A_2\,\alpha_1^*\,B_1\,\alpha_1+A_1\,\alpha_1\,B_2\,\eta^*+A_1\,\eta\,\alpha_1^*+B_1\,\eta\,\eta^*+\\
&\qquad\quad+\sum_{s=2}^{r-1}\big(A_{s+1}\,\alpha^*_{s}\,B_{s}\,\alpha_s-B_{s+1}\,\alpha^*_{s}\,A_{s}\,\alpha_s\big)\end{aligned}\\
&\label{crW}\begin{aligned}\cw(C_r)&=A_1\,\alpha_1^*\,B_2\,\alpha_1+A_2\,\alpha_1\,B_1\,\eta^*+A_2\,\eta\,\alpha_1^*+B_2\,\eta\,\eta^*+\\
&\qquad\quad+\sum_{s=2}^{r-1}\big(A_{s}\,\alpha^*_{s}\,B_{s+1}\,\alpha_s-B_{s}\,\alpha^*_{s}\,A_{s+1}\,\alpha_s\big)\end{aligned}\\
&\label{g2g2g2}\begin{aligned}\cw(G_2)&=A_1\,\alpha^*\,B_2\,\alpha- A_2\,\alpha\,B_1\,\xi^*-A_2\,\xi\,\eta^*-A_2\,\eta\,\alpha^*+B_2\,\xi\,\xi^*+B_2\,\eta\,\eta^*\end{aligned}\\
&\begin{aligned}\cw(F_4)&=A_1\,\alpha_1^*\,B_2\,\alpha_1-
B_1\,\alpha_1^*\,A_2\,\alpha_1+
A_2\,\alpha_2^*\,B_3\,\alpha_2+
A_3\,\alpha_2\,B_2\,\eta^*+\\
&\qquad+A_3\,\eta\,\alpha_2^*+B_3\,\eta\,\eta^*+
A_3\,\alpha_3^*\,B_4\,\alpha_3-
B_3\,\alpha_3^*\,A_4\,\alpha_3.
\end{aligned}
\end{align}

The only thing that remains to be done is to check the consistency of the reduction
\begin{equation}
\mathsf{rep}(Q,\cw)\rightarrow \mathsf{rep}(Q^\prime,\cw^\prime)\equiv \mathscr{L}.
\end{equation}
This requires two properties to hold. First the non--perturbative category $\mathsf{rep}(Q,\cw)$ should be consistent with the perturbative Higgs mechanism \cite{cattoy,half} with requires
\begin{equation}\label{poik}
X\in \mathscr{L}\quad \Rightarrow\quad X\big|_{\mathbf{Kr}_i}\in \mathscr{L}(SU(2)).
\end{equation}
This property, common to all quivers of a $\cn=2$ gauge theory, implies the quiver reduction $Q\rightarrow Q^\prime$ for $\mathscr{L}$.
The proof of  \eqref{poik} is by now standard \cite{cattoy,half}  
and it is spelled out for the $C_2$ example in appendix \ref{app:lemma}.
Then one has to show that  all representations of $Q^\prime$ which satisfy the Jacobian relations $\partial\cw^\prime=0$ also satisfy the original relations $\partial\cw=0$ with the $B_i$'s set to $1$. In other words, one has to check that the relations
\begin{equation}\label{uuuuq9}
\partial_{B_i}\cw\big|_{B_i=1}=0
\end{equation}
are automatically satisfied. The check is presented in appendix \ref{appendix}.
\bigskip

$\lambda=\infty$ corresponds to setting $A_i=1$, $B_i=0$. E.g.\! for $B_2$ we reduce to the representations of the quiver
\begin{equation}
\begin{gathered}
\xymatrix@R=2.0pc@C=3.0pc{
&1 \ar@<-1.4ex>@/_1.3pc/[rrr]_{\eta}  \ar@<-0.7ex>[rrr]_{\alpha} &&&\ar@<-0.7ex>[lll]_{\alpha^*}  2\ar@<-1.4ex>@/_1.3pc/[lll]_{\eta^*} 
}
\end{gathered}
\end{equation}
subjected to the relations
\begin{equation}
\eta^*\alpha_1=\eta=\alpha_1^*=0.
\end{equation}
Deleting the zero arrows, we get a string algebra whose strings have, again, dimensions in $\Delta^+(B_2)$.

\section{Other applications of the monodronic construction}

The monodromic quiver construction has also other applications besides the quivers for $\cn=2$ super--Yang Mills with non--simply laced gauge groups. We stress that most considerations in this section are \emph{purely formal} (they may or may \emph{not} correspond to consistent QFTs), and should be taken with a grain of salt.
\bigskip

The methods of refs.\!\!\cite{ACCERV2,cattoy}
give the pair $(Q,\cw)$ for $G=ADE$ SQCD with quarks in representations which are directs sums of the irrepr.\!\!  $F_i$ whose
Dynkin labels have the form
$$F_i\equiv [0,0,\cdots, 0, 1,0,\cdots, 0],\qquad 1\ \text{in the $i$--th position, }i=1,2,\cdots, r $$
 (these $r$ irrepr.\! are called \emph{fundamental} representations of $G$). At the formal level, the methods of \cite{ACCERV2,cattoy}may be extended to \emph{reducible} representations of the form \cite{ACCERV2}
\begin{equation}\label{tensorpro}
\overbrace{F_i\otimes F_i\otimes F_i \otimes \cdots \otimes F_i}^{n\ \ \text{times}},
\end{equation}
and their direct sums.
For instance, the usual quiver for $SU(2)$ $\cn=2^*$ corresponds in facts to $SU(2)$ SYM coupled to a hypermultiplet in the reducible representation\cite{ACCERV2,cattoy}
\begin{equation}
\mathbf{2}\otimes\mathbf{2}= \mathbf{3}\oplus \mathbf{1}.
\end{equation}
Indeed, the corresponding light subcategory contains an extra singlet hypermultiplet.
\smallskip

It would be desirable to have a more general construction which allows to construct $(Q,\cw)$ for all matter representations and, in particular, for the symmetric products
\begin{equation}
\mathrm{Sym}^nF_i\equiv [0,\dots, 0,n,0,\dots, 0].
\end{equation} 

The monodromic construction allows one to construct natural pairs $(Q,\cw)$ whose light subcategories contain precisely the $ADE$ $W$--bosons plus a hyper in the $\mathrm{Sym}^nF_i$. Basically, making $\lambda\rightarrow e^{2\pi i}\lambda$, induces a cyclic permutation of the direct factors in eqn.\eqref{tensorpro}. (Of course, the matter should be located in an Abelian subcategory
$\mathscr{L}_{\lambda_0}$ over a point $\lambda_0\in\mathbb{P}^1$ which is not a fixed point of the $U(1)$ action, that is, $\lambda_0\neq 0,\infty$).
\medskip

There is a special reason to be interested in such extensions. In \cite{cattoy} it was noticed the following remarkable fact: \textit{$\cn=2$ SYM with an $ADE$ gauge group $G$ coupled to a hypermultiplet in the $F_i$ is \emph{asymptotically free} if and only if the \emph{augmented} graph $G[i]$ (obtained by adding a node to the Dynkin graph of $G$ connected by a single edge to the $i$--th node) is again an $ADE$ Dynkin graph.} This does not reproduce the full list of asymptotically--free $ADE$ SQCDs because there are matter representations not of the form $F_i$ which are consistent with asymptotic freedom; in facts, for $\mathfrak{g}=ADE$ there is just one infinite series of such non--fundamental asymptotically--free representations, namely $SU(N)$ coupled to the symmetric two--index representation of dimension $\mathbf{(N+1)N/2}$. (This representation is not in the list of the `nice' ones admitting a Type IIB geometric engeneering \cite{Tack}.). In \cite{cattoy} it was suggested that, maybe, this last case is also consistent with the correspondence 
$$\text{(augmented Dynkin graphs)}\ \longleftrightarrow\ \text{(asymptotically free $\cn=2$ SQCD)}  $$
provided one identifies the relevant augmented Dynkin graph with the $C_N$ one. This proposal was not explored in \cite{cattoy} since, at the time, the non--symply case was not yet understood.
\smallskip

Now we are in a position to answer that question. Before doing that, however, it is better to introduce yet another `geometric' tool to construct $(Q,\cw)$ pairs for 4d $\cn=2$ theories, namely quiver \emph{specialization}.

\subsection{\emph{Specialization}: new pairs $(Q,\cw)$ from old ones}

Our construction of the $BCFG$ SYM quivers was modeled on the BIKMSV construction of non--simply laced gauge symmetries in string/$F$--theory \cite{BIKMSV}. That construction has been extended by Katz and Vafa to cover the matter sector
\cite{matterfromgeo}. In $F$--theory the gauge and matter sectors arise from fibrations of the same kind of singularities, except that matter is related to singularities in codimension one relatively to the ones responsible for the gauge interaction. In order to construct $(Q,\cw)$ for various matter contents, we mimic this `matter from geometry' idea trough the procedure we shall call \emph{specialization}. 

\subsubsection{The `heavy quark procedure' revisited}\label{heavyquark}

We start with an alternative viewpoint on the `heavy quark procedure' \cite{CNV,ACCERV2,cattoy,half} for the construction of the quiver for $\cn=2$ SYM with gauge group $G=ADE$ coupled to a hypermultiplet in a fundamental representation $F_i$ of $G$.
One starts by constructing the augmented graph $G[i]$ and defines its `Cartan' matrix as the $(r+1)\times (r+1)$ matrix $C[i]\equiv 2-I$, $I$ being the incidence matrix of the augmented graph $G[i]$. Then one considers the \textit{fake} SYM quiver with exchange matrix
\begin{equation}\label{ccccrrtt}
B=C[i]\otimes S,
\end{equation}
equipped with the standard square--tensor--product $G[i]\,\square\,\widehat{A}(1,1)$ superpotential \cite{kellerP}\!\!\cite{cattoy}
\begin{equation}\label{ccccrrtt2}
\cw_{G[i]\,\square\,\widehat{A}(1,1)}=\sum_{\alpha\ \text{link}\atop \text{of }G[i]}\Big(A_{t(\alpha)}\,\alpha \,B_{s(\alpha)}\,\alpha^*-B_{t(\alpha)}\,\alpha\, A_{s(\alpha)}\,\alpha^*\Big)
\end{equation}
Notice that \eqref{ccccrrtt}\eqref{ccccrrtt2} is not, in general, an ADE SYM quiver pair, since $G[i]$ needs not to be a Dynkin graph, and therefore $C[i]$ is not necessarily a Cartan matrix of  $ADE$ Lie algebra. According to \cite{cattoy}, $B$ would be a \emph{bona fide} ADE SYM quiver if and only if the coupled model is asymptotically free. 

The pair $(Q,\cw)$ for $G$ SYM coupled to a hypermultiplet in the representation $F_i$ is obtained from the above fake SYM one by setting the two arrows $A_0, B_0$ of the vertical Kronecker subquiver over the augmentation node 0 of $G[i]$ equal to two fixed complex numbers $(a_0,b_0)\equiv \lambda_0\in\mathbb{P}^1$.
The sink and source nodes of $\mathbf{Kr}_0$ are then identified \emph{via} the isomorphism $B_0$,
\begin{equation}
\begin{gathered}
\xymatrix{& \ar@{-->}[l] \bullet\ar@{-->}[r]&\\
\\
 \ar@{-->}[r] & \bullet\ar@<0.4ex>[uu]\ar@<-0.4ex>[uu]   &\ar@{-->}[l]}\\
 \mathbf{Kr}_0
\end{gathered}\quad \longrightarrow\quad 
\begin{gathered}
\xymatrix{ && \\
& \bullet\ar@{-->}[ul]\ar@{-->}[ur]\\
 \ar@{-->}[ru] &  & \ar@{-->}[lu]}\\
f
\end{gathered}
\end{equation}
so that the resulting quiver has $2r+1$ nodes and the exchange matrix has one zero eigenvalue, corresponding to the flavor symmetry $f$ of the matter sector. The associated superpotential is
\begin{equation}
\cw= \cw_{G[i]\,\square\,\widehat{A}(1,1)}\Big|_{(A_0,B_0)=(\lambda_0,1)}.
\end{equation}

 Thus the SQCD pair $(Q,\cw)$ is just a \textit{specialization} at the augmented node of the fake SYM pair. This implies that the generic light category $\mathscr{L}_\lambda$ is equivalent to the pure $G$ SYM one, while the \emph{specialized} category $\mathscr{L}_{(a_0:b_0)}$ coincides with the one for `pure $G[i]$ fake SYM'. Of course the last category is well behaved iff $G[i]$ is a Dynkin quiver \cite{cattoy}. (Physically, specialization corresponds to replacing the would be simple--root $W$--boson vector associated to the augmentation node with a scalar).

\subsubsection{More general (split) specializations: quiver gauge theories}

In \S.\,\ref{heavyquark} we specialized the augmented square--tensor--products of the form $G[i]\,\square\,\widehat{A}(1,1)$ at the augmentation Kronecker subquiver $\upuparrows$. We could as well have specialized at any other Kronecker subquiver, or even at several such subquivers.
\smallskip

A first generalization of the `heavy quark procedure' may be described as follows. Let $P$ be any alternating quiver (recall that a quiver is \emph{alternating} if all its nodes are either sinks or sources; all connected trees admit two alternating orientations). 
Then the square--tensor--product $P\,\square\,\widehat{A}(1,1)$ is  a well defined \cite{kellerP} pair $(Q_P,\cw_P)$ with a vertical Kronecker subquiver for each node in $P$. If $I\subset P_0$ is a subset of the nodes of $P$, the specialization at $I$ corresponds to specializing
\begin{equation}
(A_i,B_i)\rightsquigarrow (a_i,b_i)\in \mathbb{P}^1\quad \text{for all }i\in I,
\end{equation}
while identifying in pairs the sources and sinks of the Kronecker subquivers $\mathbf{Kr}_i$ for $i\in I$.

For instance, we can specialize $A_N\,\square\,\widehat{A}(1,1)$ at the $K$--th Kronecher subquiver. The resulting pair $(Q,\cw)$ corresponds to the quiver gauge theory with gauge quiver
\begin{equation}
\begin{gathered}
\xymatrix{*++[o][F-]{\phantom{K}}\ar[rr] && *++[o][F-]{\phantom{K}}}
\end{gathered}
\end{equation}
where the nodes stand, respectively, for $SU(K)$ and $SU(N-K)$ gauge groups and the arrow to a $(N-K,\overline{K})$ bifundamental hypermultiplet.
\smallskip

From this example it is pretty obvious that we may construct the pair $(Q,\cw)$ of any quiver (in the gauge sense) $\cn=2$ model with $ADE$ gauge groups simply by specializing a suitable $P\,\square\,\widehat{A}(1,1)$ at the appropriate set $I$ of Kronecker subquivers. \smallskip

The  specializations of $ADE$ SYM at \emph{one} Kronecker precisely correspond to the list of cases explicitly discussed in ref.\!\cite{matterfromgeo}, namely
\begin{equation}\label{katz}
\begin{gathered}
 A_n\rightarrow A_{n-k}\times A_{k-1}\\
D_n\rightarrow D_{n-1}, A_{n-1}, D_{n-r}\times A_{r-1}\\
E_6\rightarrow D_5, A_5\\
E_7\rightarrow D_6,E_6,A_6\\
E_8\rightarrow E_7.
\end{gathered}
\end{equation}
In all instances the matter representations found in \cite{matterfromgeo} using geometric methods agree with those found in \cite{cattoy} using the Representation Theory of the associated quivers with superpotential. In facts, in both approaches one is effectively reduced to a breaking of the adjoint representation of the Lie groups in the left part of eqn.\eqref{katz} to the Lie groups on the right \emph{times} $U(1)_f$, where --- after specialization --- $U(1)_f$ is interpreted as the \emph{global} flavor symmetry of the matter sector. (Note that specialization at $k$ vertical Kronecker subquivers reduces the numbers of nodes of the quiver by $k$; in particular, the exchange matrices of one-Kronecker specializations have odd rank, so that $\det B=0$, consistent with the fact that there is a flavor $U(1)_f$ charge). 

\medskip

For more general 4d $\cn=2$ gauge models we have to combine specialization with the monodronic constructions discussed in the first part of the present paper.

\subsection{Coupling matter in  $\mathrm{Sym}^n\, F_i$}\label{symsymsym}

The above construction by specialization may be generalized to the case in which the link between the augmentation node 0 and the $i$--th node of $G$ has a non trivial value $(1,n)$, with $n\in \mathbb{N}$. We claim that, after the specialization $$(A_0,B_0)=(a_0,b_0)\in \mathbb{P}^1,$$ this construction corresponds to coupling to $G$ SYM a hypermultiplet in the representation $\mathrm{Sym}^n\,F_i$ of $G$.
\medskip

The physical case of this construction is $G=A_{N-1}$ coupled to
$\mathrm{Sym}^2\,\mathbf{N}=\mathbf{(N+1)N/2}$. The corresponding augmented valued graph is the
$C_N$ Dynkin diagram. The generic category $\mathscr{L}_\lambda$ has bricks with charges equal to the positive roots of $SU(N)$, while the bricks of the specialized  category $\mathscr{L}_{\lambda_0}$ have charges corresponding to the positive roots of $C_N$. Since
\begin{equation}
\mathrm{Ad}\,Sp(N)=\mathrm{Ad}\, SU(N)\oplus \mathbf{(N+1)N/2}\oplus\overline{\mathbf{(N+1)N/2}}\oplus \boldsymbol{1},
\end{equation}
we get the claim for this example. In particular, this shows that the conjecture of \cite{cattoy} is true.
\smallskip

The Langlands dual construction in which one interchanges the role of long and short roots in the augmented valued graph, with the effect of replacing $C_N$ with $B_N$, corresponds to a non--generic (in the sense of \S.\,8.4 of \cite{cattoy}) construction  of the pair $(Q,\cw)$ for the coupling to $SU(N)$ SYM to a hypermultiplet in the reducible representation
\begin{equation}
\mathbf{N}\oplus \mathbf{N(N-1)/N}.
\end{equation}
in agreement with the group--theoretical branch rule
\begin{equation}
\mathrm{Ad}\,SO(2N+1)=\mathrm{Ad}\, SU(N)\oplus \mathbf{N(N-1)/2}\oplus\overline{\mathbf{N(N-1)/2}}\oplus \mathbf{N}\oplus\overline{\mathbf{N}}\oplus\boldsymbol{1},
\end{equation}

The specialization of $G_2$ SYM leads to the quiver for $SU(2)$ coupled to a hypermultiplet of isospin $3/2$ already discussed in \cite{CNV} (see also \cite{cattoy}). The corresponding QFT is not UV complete, and, while the light subcategory $\mathscr{L}$ is perfectly nice, the full non--perturbative category $\mathsf{rep}(Q,\cw)$ is expected to be quite wild \cite{cattoy} (it \emph{is} wild in the RT technical sense).

\section{Coupling $BCFG$ SYM to matter}

A detailed analysis of all models arising from the coupling of $BCFG$ $\cn=2$ SYM to matter is beyond the scope of the present paper. Here we limit ourselves to a sketchy discussion of some simple models, leaving a more systematic treatment to future work.\medskip 

We focus on the non--simply laced analogues of the `matter from geometry' models in eqn.\eqref{katz}, namely one--Kronecker specializations of $BCFG$ SYM quivers (different from the ones corresponding to $ADE$ SYM coupled to $\mathrm{Sym}^n\,F_i$ matter already discussed in \S.\,\ref{symsymsym}). The non--simply laced counterpart to eqn.\eqref{katz} is\footnote{\ We omit the gauge singlets.}
\begin{gather}\label{nonkatz}
\begin{array}{c|c}
\phantom{\Big|}\mathfrak{g}_\text{geometry}\rightarrow \mathfrak{g}_\text{gauge} & \text{matter representations} \\\hline
B_n\rightarrow A_k\times B_{n-k-1}  & \big(\mathbf{k+1},\ \mathbf{2(n-k)-1}\big) \\
C_n\rightarrow A_k\times C_{n-k-1} & \big(\mathbf{k+1},\ \mathbf{2(n-k-1)}\big) \\
F_4\rightarrow  B_3  &\mathbf{8}\oplus\mathbf{7} \\
F_4\rightarrow  C_3  & \mathbf{14}\equiv [0,0,1] \\\hline
\end{array}
\end{gather}
where the matter consists of one hypermultiplet in the indicated representation of $\mathfrak{g}_\text{gauge}$.  The quivers and superpotentials for the various models are obtained straightforwardly from the specialization procedure. For instance, in the
$A_k\times B_{n-k-1}$ case, the quiver is

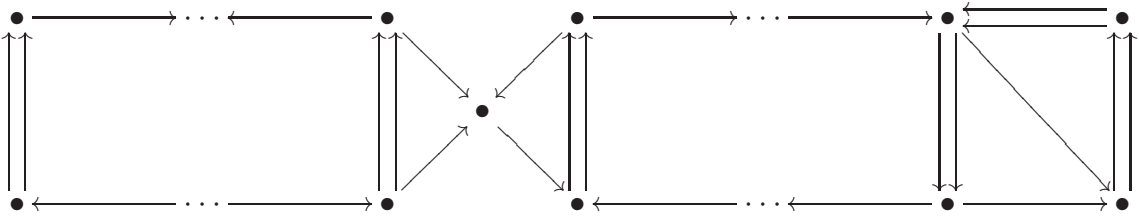
\begin{equation*}
 \begin{gathered}
 \xymatrix{\bullet \ar[rr] && \cdots &&\bullet\ar[ll]\ar[rd] && \bullet\ar[ld] \ar[rr] && \cdots \ar[rr] && \bullet\ar@<0.6ex>[dd]\ar@<-0.6ex>[dd]\ar[ddrr] &&\bullet \ar@<0.6ex>[ll]\ar@<-0.6ex>[ll]\\
 && && &\bullet\ar[dr]\\
\bullet\ar@<0.6ex>[uu]\ar@<-0.6ex>[uu] &&\cdots\ar[ll] \ar[rr] &&\bullet \ar@<0.6ex>[uu]\ar@<-0.6ex>[uu]\ar[ur] && \bullet\ar@<0.6ex>[uu]\ar@<-0.6ex>[uu]&& \cdots\ar[ll] && \bullet\ar[ll] \ar[rr] &&\bullet \ar@<0.6ex>[uu]\ar@<-0.6ex>[uu]} 
 \end{gathered}
\end{equation*}
where the vertical Kronecker subquiver $\mathbf{Kr}_{n-k}$ in the first quiver of figure \ref{nonquivers} has being contracted to a single node atteched by oriented triangles \cite{CV11} to the rest of the quiver. Correspondingly, 
the superpotential is simply a specialization of the pure $SO(2n+1)$ SYM one
\begin{equation}
 \cw_{SU(k+1)\times SO(2n-2k-1)}=\cw_{SO(2n+1)}\Big|_{(A_{n-k},B_{n-k})=(\lambda_0,1)}.
\end{equation}

\medskip
 As explained in \cite{ACCERV2,cattoy}, if we have $N_f$ flavors, to get the pair $(Q,\cw)$ we just reiterate the same construction $N_f$ times.\smallskip

In particular, taking $k=0$ in the first two rows of eqn.\eqref{nonkatz}, we get the pairs $(Q,\cw)$ for $SO(2m+1)$ and $Sp(m)$ SYM coupled to $N_f$ fundamentals.

\section*{Acknowledgements}

We have greatly benefited of discussions with
Murad Alim,  Clay C\'ordova, Sam Espahbodi,
Ashwin Rastogi and especially Cumrun Vafa.  We thank them all.
SC thanks the Simons Center for Geometry and Physics, where this work was completed, for hospitality.
\newpage
\appendix

\section{Consistency with the perturbative Higgs mechanism}\label{app:lemma}

The consistency of our pairs $(Q,\cw)$ with the perturbative Higgs mechanism requires the following mathematical \textbf{Fact} to hold
\medskip

{\bf Fact.} \textit{If $X \in \mathscr{L}$, then
\begin{equation}
X\Big|_{\mathbf{Kr}_i} \in \mathscr{L}(SU(2))\equiv\mathcal{T}_\mathbf{Kr}\qquad \forall\, i.
\end{equation}}
\medskip

Here $\mathcal{T}_\mathbf{Kr}$ stands for the Abelian category of the regular representations of the Kronecker quiver \cite{RI,ASS,CB}. The strategy of the proof of this kind of results is rather standard see the appendices of ref.\!\cite{cattoy,half}.
Here we limit ourselves to present the details for the $B_2$ example.

\medskip

\textsc{Proof} (for $G=B_2$).  The $B_2$ quiver and superpotential
are \begin{equation}
\begin{gathered}
\xymatrix{ 2,1 \ar@<0.5ex>@{->}[ddd]^{A_2}\ar@<-0.5ex>@{->}[ddd]_{B_2} \ar[dddrrr]^{\eta}&&& 1,2 \ar@<0.5ex>@{->}[lll]^{\eta^*}\ar@<-0.5ex>@{->}[lll]_{\a^*}\\
&&&\\
&&&\\
2,2 \ar[rrr]^{\a}&&& 1,1\ar@<0.5ex>@{->}[uuu]^{A_1}\ar@<-0.5ex>@{->}[uuu]_{B_1}}
\end{gathered}
\end{equation}
$$ \cw= A_2 \a^* B_1 \a + A_1 \a B_2 \eta^* + A_1 \eta \a^* + B_1 \eta \eta^* .$$
Since $m_i(X)=0$, if $X |_\mathbf{Kr} \notin \mathcal{T}$ it must have a preinjective summand. Consider first $\mathbf{Kr}_1$. If $X|_{\mathbf{Kr}_1}$ has a preinjective summand, there are vectors  --- not \emph{all} zero (\cite{cattoy} appendix) --- $v_1,\dots,v_{\ell} \in X_{1,1}$ such that
\begin{equation}\label{injsummd}
A_1 v_1 = 0,\qquad A_1 v_{a+1} = B_1 v_a, \text{ for } a = 1,..., \ell-1 \qquad B_1 v_\ell = 0.
\end{equation}
Let us prove by induction on $\ell$ that, if this is the case, then $X \notin \mathscr{L}(B_2)$. The case $\ell =1$ is trivial: if $A_1 v = 0 = B_1 v$, then $\C v$ is subrepresentation with positive $m_1$ magnetic charge and is not in $\mathscr{L}$. Assuming that it is true for $\ell -1$ vectors, let us now prove that it is true also for $\ell$ of them. By the two equations
\begin{gather}\label{sgrbvz1}
\partial_{\eta} \cw = \a^* A_1 + \eta^* B_1 = 0 \\
\partial_{\a} \cw = A_2 \a^* B_1 + B_2 \eta^* A_1 = 0, \label{sgrbvz2}
\end{gather}
we have that there exist $\ell - 1$ vectors
\begin{equation}
\framebox[1.2\width]{$w_i \equiv \eta^* B_1 v_{i+1} $} \in X_{2,1} \qquad i = 1 , \dots , \ell-1
\end{equation}
that are such that $w_0 = 0$, $w_{\ell -1} = 0$, and
\begin{equation}\label{wwprty}
\begin{aligned}
A_2 \ w_a &\equiv A_2 \eta^* B_1 \ v_{a+1}\\
&=_\text{eqn.\eqref{sgrbvz1} } - A_2 \a^* A_1 \ v_{a+1} =\\
&=_\text{ eqn.\eqref{injsummd} } - A_2 \a^* B_1 \ v_a =\\
&=_\text{eqn.\eqref{sgrbvz2} }  B_2 \eta^* A_1 \ v_a =\\
&=_\text{ eqn.\eqref{injsummd} } B_2 \eta^* B_1 \ v_{a-1} =\\
&= B_2 \ w_{a-2}.
\end{aligned}
\end{equation}
In particular, this implies that
\begin{equation}
A_2 \ w_1=  A_2 \ w_2 = 0 =  B_2 \ w_{\ell - 2}.
\end{equation}
Notice that if the vectors $w_i$ are all equal to zero, then we have that $X|_{\mathbf{Kr}_1}$ is a subrepresentation of $X$ that being a preinjective summand of $X|_{\mathbf{Kr}_1}$, has positive $m_1$ charge. Now, let us define
\begin{equation}
\framebox[1.2\width]{$z_i \equiv \eta w_i + \a B_2 w_{i-1}$} \in X_{1,1} \qquad i = 1, \dots, \ell -1.
\end{equation}
If all the $z_a$ are equal to zero, then by \eqref{wwprty},
\begin{equation}\label{zetallzero}
\eta \ w_a = - \a B_2 \ w_{a-1} = - \a A_2 \ w_{a+1}.
\end{equation}
Then, if also $\eta \ w_a \equiv 0$, eqn.\eqref{sgrbvz1}\eqref{sgrbvz2}, and
\begin{equation}\label{sgrbvz3}
\partial_{A_1} \cw = \a B_2 \eta^* + \eta \a^* = 0
\end{equation}
give the additional relations
\begin{equation}\label{frufru}
\begin{aligned}
0  &\equiv \eta \ w_a \equiv \eta \eta^* B_1 \ v_{a+1} =_\text{ eqn.\eqref{sgrbvz1} } - \eta \a^* A_1\ v_{a+1} \\
&=_\text{eqn.\eqref{sgrbvz3} } \a B_2 \eta^* A_1\ v_{a+1} 
=_\text{ eqn.\eqref{sgrbvz2} } - \a A_2 \a^* B_1\ v_{a+1}.  
\end{aligned}
\end{equation}
Now, for each node of $Q(B_2)$, consider the following subspaces:
\begin{equation}
\begin{aligned}
& V_{1,1} \equiv \text{span}(v_1,...,v_{\ell}) \subset X_{1,1} \\
& V_{1,2} \equiv B_1 V_{1,1} = A_1 V_{1,1} \subset X_{1,2} \\
& V_{2,1} \equiv \eta^* V_{1,2} = - \a^*V_{1,2} \subset X_{2,1}\\
& V_{2,2} \equiv A_2 V_{2,1} = B_2 V_{2,1} \subset X_{2,2}.
\end{aligned}
\end{equation}
The relations we have just obtained in \eqref{frufru} entail that
\begin{equation}
\eta \ V_{2,1} \equiv 0 \quad\text{and}\quad \a \ V_{2,2} \equiv 0,
\end{equation}
therefore, if $\eta \ w_a \equiv 0$, the module $X$ has the following submodule (arrows are defined by restriction)
\begin{equation}
V\equiv (V_{1,1},V_{1,2},V_{2,1},V_{2,2}) \subset X.
\end{equation}
Since
\begin{equation}
V|_{\mathbf{Kr_1}}  \ \ \text{is preinjective} \Rightarrow m_1 (V) > 0.
\end{equation}
So, we have shown that $X$ has a submodule with positive $m_1$ magnetic charge:  This proves that if $X$ is such that $\eta \ w_a \equiv 0$, $X \notin \mathscr{L}(B_2)$. Thus, assume that $\eta \ w_a$ are not all zero. If this is the case, but we still have that $z_a \equiv 0$, composing $B_1$ to \eqref{zetallzero} on the left, we have that
\begin{equation}
B_1 \eta \ w_a = - B_1 \a A_2 \ w_{a+1},
\end{equation}
but by
\begin{equation}\label{fru1}
\partial_{\a^*} \cw = B_1 \a A_2 + A_1 \eta =0
\end{equation}
we have that
\begin{equation}
A_1 \eta \ w_a = -B_1 \a A_2 \ w_a
\end{equation}
Thus
\begin{equation}\label{etaww1}
A_1 \eta \ w_a = B_1 \eta \ w_{a-1}.
\end{equation}
Moreover, by \eqref{injsummd}, \eqref{sgrbvz1}, and \eqref{sgrbvz3} we obtain
\begin{equation}\label{etaww2}
\begin{aligned}
A_1 \eta \ w_1&\equiv A_1 \eta \eta^* B_1 \ v_2\\
& =_\text{eqn.\eqref{sgrbvz1} } - A_1 \eta \a^* A_1 \ v_2 \\
&=_\text{ eqn.\eqref{injsummd}} - A_1 \eta \a^* B_1\ v_1\\
&=_\text{ eqn.\eqref{sgrbvz3}} A_1 \a B_2 \eta^* B_1 \ v_1\\
&=_\text{ eqn.\eqref{sgrbvz1} }  - A_1 \a B_2 \a^* A_1 \ v_1\\
&= 0.
\end{aligned}
\end{equation}
That, together with
\begin{equation}
B_1 \eta \ w_{\ell -1} = 0,
\end{equation}
that holds by definition, entail that there are $\ell -1$ vectors (not all zero), namely
\begin{equation}
\eta \ w_a \in X_{1,1}\qquad a = 1,...,\ell-1,
\end{equation}
such that \eqref{injsummd} holds: Thus by the inductive hypothesis, $X \notin \mathscr{L}(B_2)$. Therefore, we can safely assume that the $z_a$'s are not all zero: By the equations of motion \eqref{fru1} and
\begin{equation}\label{fru2}
\partial_{\eta^*} \cw = A_1 \a B_2 + B_1 \eta = 0
\end{equation}
the vectors $z_a$ are such that, for $a = 2,..., \ell-1$
\begin{equation}
\begin{aligned}
A_1 \ z_a &\equiv A_1 \eta \ w_a + A_1 \a B_2 \ w_{a-1}\\
&=_\text{ eqns.\eqref{fru1},\eqref{fru2} } - B_1 \a A_2 \ w_a - B_1 \eta \ w_{a-1}\\
&=_\text{ eqn.\eqref{wwprty} } - B_1 \a B_2 \ w_{a-2} - B_1 \eta \ w_{a-1} \\
&= - B_1 \ z_{a-1}.
\end{aligned}
\end{equation}
Then, since $w_0 = 0$,
\begin{equation}
A_1 \ z_1 \equiv A_1 \eta \ w_1 =_\text{ eqn.\eqref{etaww2}} = 0
\end{equation}
and, since $w_{\ell-1}=0$,
\begin{equation}
\begin{aligned}
B_1 z_{\ell-1} &\equiv \a B_2 w_{\ell -2}\\
&\equiv \a B_2 \eta^* B_1 v_{\ell-1}\\
&=_\text{eqn.\eqref{injsummd}} \a B_2 \eta^* A_1 v_{\ell}\\
&=_\text{eqn.\eqref{sgrbvz3}} - \eta \a^* A_1 v_{\ell}\\
&=_\text{eqn.\eqref{sgrbvz1}} \eta \eta^* B_1 v_{\ell}\\
&= 0.
\end{aligned}
\end{equation}
So we have shown that if there is a family of $\ell$ vectors of type \eqref{injsummd}, then either $X$ has a positive magnetic--charge submodule or there is an analogous family of $\ell -1$ vectors (not all zero), and therefore $X|_{\mathbf{Kr}_1}$ cannot have preinjective summands, by induction, and it must be regular. Now, consider $X|_{\mathbf{Kr}_2}$. If $X|_{\mathbf{Kr}_2}$ has a preinjective summand then there are $v_1,...,v_\ell$ vectors in $X_{2,1}$ such that
\begin{equation}\label{injsummd2}
A_2 v_1 = 0,\qquad A_2 v_{a+1} = B_2 v_a, \text{ for } a = 1,..., \ell-1 \qquad B_2 v_\ell = 0.
\end{equation}
Again we will proceed by induction on $\ell$. The case $\ell =1$ is trivial: we have $A_2 v = 0 = B_2 v$ and so that by eqns.\eqref{fru1},\eqref{fru2}
\begin{equation}
A_1 \eta \ v= 0 =  B_1 \eta \ v.
\end{equation}
If $\eta v = 0$ the $\C v \subset X_{2,1}$ is a destabilizing subrep, if not, then $\C \ \eta v \subset X_{1,1}$ is one. Now define
\begin{equation}
\framebox[1.2\width]{$u_a \equiv \eta^*B_1\a B_2 \ v_a$} \in X_{2,1} \qquad i = 1, \dots, \ell -1.
\end{equation}
Now, if the $u_a$ are zero, then, by
\begin{equation}\label{fru3}
\partial_{\eta^*} \cw = A_1 \a B_2 + B_1 \eta = 0
\end{equation}
we have the relations
\begin{equation}
\begin{aligned}
0&=\eta^*A_1 \eta \ v_{a+1}\\
&=_\text{eqn.\eqref{fru1}} \eta^* B_1 \a A_2 \ v_{a+1} \\
&=_\text{eqn.\eqref{injsummd2}} \eta^*B_1\a B_2 \ v_a\\
&=_\text{eqn.\eqref{sgrbvz1}} \a^* A_1 \a B_2 \ v_a =_\text{eqn.\eqref{injsummd2}}  \a^* A_1 \a A_2 \ v_{a+1}\\
& =_\text{eqn.\eqref{fru3}} \a^* B_1 \eta \ v_a.
\end{aligned}
\end{equation}
that, together with
\begin{equation}
\begin{aligned}
&\partial_{A_2} \cw = \a^* B_1 \a = 0\\
&\partial_{B_2} \cw = \eta^* A_1 \a = 0\\
&\eta^* B_1 \eta =_\text{eqn.\eqref{fru2}} \eta^* A_1 \a B_2 = 0\\
&\a^* A_1 \eta =_\text{eqn.\eqref{fru1}} \a^* B_1 \a A_2 = 0.
\end{aligned}
\end{equation}
entail, proceeding as above, that, the following subspaces
\begin{equation}
\begin{aligned}
&V^{\prime}_{2,1} \equiv \text{span}(v_1,...,v_{\ell}) \subset X_{2,1}\\
&V^{\prime}_{2,2} \equiv B_2 \ V^{\prime}_{2,1} = A_2 \ V^{\prime}_{2,1} \subset X_{2,2}\\
&V^{\prime}_{1,1} \equiv \a \ V^{\prime}_{2,2} + \eta \ V^{\prime}_{2,2} \subset X_{1,1}\\
&V^{\prime}_{1,2} \equiv A_1 V^{\prime}_{1,1} + B_1 V^{\prime}_{1,1} \subset X_{1,2}
\end{aligned}
\end{equation}
are such that
\begin{equation}
\a^* V^{\prime}_{1,2} = 0 = \eta^* V^{\prime}_{1,2}.
\end{equation}
Therefore, if the module $X$ is such that $u_a \equiv 0$, such a module has a submodule (arrows are defined by restriction)
\begin{equation}
V^{\prime} \equiv (V^{\prime}_{1,1},V^{\prime}_{1,2},V^{\prime}_{2,1},V^{\prime}_{2,2}) \subset X.
\end{equation}
But
\begin{equation}
V^{\prime}|_{\mathbf{Kr_2}}  \ \ \text{is preinjective} \Rightarrow m_2 (V^{\prime}) > 0,
\end{equation}
therefore $X$ has a submodule of positive magnetic charge and it cannot belong to the light subcategory. This shows that we can assume that the $u_a$ are not all zero.  Now, by \eqref{sgrbvz1},\eqref{sgrbvz2}, \eqref{fru1}, \eqref{fru2}, and \eqref{injsummd2}
\begin{equation}
\begin{aligned}
A_2 \ u_a &\equiv A_2 \eta^* B_1 \a B_2 v_a\\
&=_\text{eqn.\eqref{sgrbvz1}}- A_2 \a^* A_1 \a B_2 v_a \\
&=_\text{eqn.\eqref{fru2}} A_2 \a^* B_1 \eta v_a\\
&=_\text{eqn.\eqref{sgrbvz2}} B_2 \eta^* A_1 \eta v_a \\
&=_\text{eqn.\eqref{fru1}} B_2 \eta^* B_1 \a A_2 v_a\\
&=_\text{eqn.\eqref{injsummd2}} B_2 \eta^* B_1 \a B_2 v_{a-1}\\
&= B_2 \ u_{a-1}.
\end{aligned}
\end{equation}
This also shows that $A_2 u_1 = 0$ and that $B_2 u_{\ell-1} =0$. So, by the inductive hypothesis on $\mathbf{Kr}_2$, $X \notin \mathscr{L}(B_2)$. $\square$

\section{Consistency of $\mathsf{rep}(Q,\cw)\rightarrow \mathsf{rep}(Q^\prime,\cw^\prime)$}\label{appendix}

We need to check eqn.\eqref{uuuuq9} only at the Kronecker subquivers associated to nodes of the Dynkin graph which are connected by a link with a non--trivial value since at all other nodes we reduce to the corresponding statement for $ADE$ SYM. In other words, we need to check only the two cases $C_2$ and $G_2$.

\subsection{$C_2$}

Let us recall $(Q,\cw)$ for $C_2$ SYM
\begin{equation}
\begin{gathered}
\xymatrix{\bullet\ar[dddrrr]^{\eta}\ar@<0.7ex>[ddd]^{B_1}\ar@<-0.7ex>[ddd]_{A_1} 
&&& \bullet\ar@<0.7ex>[lll]^{\eta^*}\ar@<-0.7ex>[lll]_{\alpha^*} \\
\\
\\
\bullet\ar[rrr]_{\alpha} &&& \bullet\ar@<0.7ex>[uuu]^{A_2}\ar@<-0.7ex>[uuu]_{B_2}}
\end{gathered}
\end{equation}
\begin{align}
\cw = 
 \alpha\, A_1\alpha^* B_2+\alpha\, B_1\, \eta^* A_2+\eta(\eta^* B_2+\alpha^* A_2)
\end{align}

Assuming the usual lemma \cite{cattoy,half} on the restriction of $X\in \mathscr{L}$ on the Kronecker subquivers to hold, we  set  $B_1,B_2=1$, and consistency of this restrictions requires the relations
\begin{gather}\label{re1}
\partial_{B_2}\cw\equiv \alpha\, A_1\alpha^*+\eta\,\eta^*=0\\
\partial_{B_1}\cw\equiv \eta^*\,  A_2\,\alpha=0\label{re2}
\end{gather}
 to hold identically for all elements of $\mathsf{rep}(Q^\prime,\cw^\prime)$ where 
\begin{align}
&Q^\prime\qquad \begin{gathered}
\xymatrix@R=2.0pc@C=3.0pc{
&\ar@(ul,dl)[]_{A_1} 1 \ar@/_1.5pc/[rrr]_{\eta} \ar@/_3.2pc/[rrr]_{\alpha} &&&\ar@/_1.5pc/[lll]_{\eta^*}\ar@/_3.2pc/[lll]_{\alpha^*}  \ar@(ur,dr)[]^{A_2} 2  
}
\end{gathered}
\end{align}
\begin{equation}
\cw^\prime\equiv\cw\big|_{B_1,B_2=1}=\alpha\, A_1\,\alpha^* +\alpha\, \eta^*\, A_2+\eta\, \eta^* +\eta\, \alpha^*\, A_2.
\end{equation}
Here we check that eqns.\eqref{re1}\eqref{re2} are indeed automatically satisfied.
As in the main body of the paper,
we may integrate out the massive fields $\eta$, $\eta^*$ using their equations of motion
\begin{gather}\label{mot1}
\eta^*=-\alpha^*\, A_2\\
\eta=-A_2\,\alpha.\label{mot2}
\end{gather}
In this way we are reduced to the category $\mathsf{rep}(Q^{\prime\prime},\cw^{\prime\prime})$ where 
\begin{align}
&Q^{\prime\prime}\qquad \begin{gathered}
\xymatrix@R=2.0pc@C=3.0pc{
&\ar@(ul,dl)[]_{A_1} 1  \ar@/_1pc/[rrr]_{\alpha} &&&\ar@/_1pc/[lll]_{\alpha^*}  \ar@(ur,dr)[]^{A_2} 2  
}
\end{gathered}
\end{align}
\begin{equation}
\cw^{\prime\prime}=\alpha\, A_1\,\alpha^* -\alpha\, \alpha^*\, A_2^2.
\end{equation}
The relations of this last Jacobian algebra are
\begin{align}
&\partial_\alpha \cw^{\prime\prime}\equiv A_1\,\alpha^*-\alpha^*\, A_2^2=0\label{rel1}\\
& \partial_{\alpha^*}\cw^{\prime\prime}\equiv \alpha\, A_1-A_2^2\,\alpha=0\label{rel2}\\
&\partial_{A_1}\cw^{\prime\prime}\equiv \alpha^*\, \alpha=0\label{rel3}\\
&\partial_{A_2}\cw^{\prime\prime}\equiv - \alpha\,\alpha^*\, A_2-A_2\,\alpha\,\alpha^*=0.\label{rel4}
\end{align}
Let us check that eqns.\eqref{re1}\eqref{re2} are consequences of these relations. Indeed, using \eqref{mot1}\eqref{mot2},
we get
\begin{multline}
\alpha\,  A_1\alpha^*+\eta\,\eta^* =\alpha A_1\alpha^*+A_2\alpha\alpha^* A_2=\\
=\alpha A_1\alpha^*-\alpha\alpha^* A_2^2+(\partial_{A_2}\cw^{\prime\prime}) A_2
=\alpha (\partial_\alpha\cw^{\prime\prime})+(\partial_{A_2}\cw^{\prime\prime}) A_2=0
\end{multline}
\begin{equation}
\eta^* A_2\alpha=-\alpha^* (A_2^2\alpha)=-\alpha^* \alpha A_1+\alpha^*(\partial_{\alpha^*}\cw^{\prime\prime})
=-(\partial_{A_1}\cw^{\prime\prime})A_1+\alpha^*(\partial_{\alpha^*}\cw^{\prime\prime})=0.\end{equation}

\subsection{$G_2$}

With reference to the $G_2$ SYM quiver in figure \ref{nonquivers}, the superpotential is given by eqn.\eqref{g2g2g2}.
Again, to restrict to the light category we set $B_1,B_2=1$ and identify the source and target nodes of the vertical Kronecker subquivers. The relations which should automatically hold for consistency are
\begin{gather}
\partial_{B_1}\cw\big|_{B_1,B_2=1}\equiv \xi^*\,A_2\,\alpha=0\label{mm1}\\
\partial_{B_2}\cw\big|_{B_1,B_2=1}\equiv\alpha\, A_1\,\alpha^*+ \xi\,\xi^*+\eta\,\eta^*=0.\label{mm2}
\end{gather}
The light category is then given in terms of $(Q^\prime,\cw^\prime)$ where
\begin{align}
&Q^\prime\qquad \begin{gathered}
\xymatrix@R=2.0pc@C=3.0pc{
&\ar@(ul,dl)[]_{A_1} 1 \ar@/_0.8pc/[rrr]_{\alpha} \ar@/_2pc/[rrr]_{\eta} 
 \ar@/_3.2pc/[rrr]_{\xi}&&&\ar@/_0.8pc/[lll]_{\alpha^*}\ar@/_2pc/[lll]_{\eta^*}\ar@/_3.2pc/[lll]_{\xi^*}   \ar@(ur,dr)[]^{A_2} 2  
}
\end{gathered}
\end{align}
\begin{equation}
\cw^\prime= \alpha\, A_1\,\alpha^* - \xi^*\,A_2\,\alpha -\eta^*\,A_2\,\xi+\xi\,\xi^*-\alpha^*\, A_2\,\eta+\eta\,\eta^*.
\end{equation}
We integrate out the massive fields $\eta, \xi, \eta^*, \xi^*$ using their equations of motion
\begin{gather}
\xi=A_2\,\alpha\label{mmmm1}\\
\xi^*=\eta^*\,A_2\\
\eta=A_2\,\xi=A_2^2\,\alpha\\
\eta^*=\alpha^*\, A_2\\
\Rightarrow\ \xi^*=\alpha^*\, A_2^2.\label{mmmm4}
\end{gather}
and we remain with the pair $(Q^{\prime\prime},\cw^{\prime\prime})$ 
\begin{align}
&Q^{\prime\prime}\qquad \begin{gathered}
\xymatrix@R=2.0pc@C=3.0pc{
&\ar@(ul,dl)[]_{A_1} 1  \ar@/_1.2pc/[rrr]_{\alpha} &&&\ar@/_1.2pc/[lll]_{\alpha^*}  \ar@(ur,dr)[]^{A_2} 2  
}
\end{gathered}
\end{align}
\begin{equation}
\cw^{\prime\prime}=\alpha\, A_1\,\alpha^* - \alpha\, \alpha^*\, A_2^3.
\end{equation}
The corresponding Jacobian relations are
\begin{align}
&\partial_\alpha \cw^{\prime\prime}\equiv A_1\,\alpha^*-\alpha^*\, A_2^3=0\label{rel13}\\
& \partial_{\alpha^*}\cw^{\prime\prime}\equiv \alpha\, A_1- A_2^3\,\alpha=0\label{rel23}\\
&\partial_{A_1}\cw^{\prime\prime}\equiv \alpha^*\, \alpha=0\label{rel3}\\
&\partial_{A_2}\cw^{\prime\prime}\equiv -(A_2^2\,\alpha\,\alpha^*+A_2\,\alpha\,\alpha^*\, A_2+\alpha\,\alpha^*\, A_2^2)=0.\label{rel43}
\end{align}
Let us check that these relations imply eqns.\eqref{mm1}\eqref{mm2}. Indeed, in view of eqns.\eqref{mmmm1}--\eqref{mmmm4},
\begin{align}
&\xi^*\, A_2\,\alpha=\alpha^*\, A_2^3\,\alpha= \alpha^*(\alpha\, A_1)=(\partial_{A_1}\cw^{\prime\prime})A_1=0,\\
&\begin{aligned}
\alpha\, A_1\,\alpha^*+&\xi\,\xi^*+\eta\,\eta^*=\alpha\, A_1\,\alpha^*+A_2\,\alpha\, \alpha^*\, A_2^2+A_2^2\,\alpha\, \alpha^*\, A_2=\\
&=\alpha\, A_1\,\alpha^* -A_2(\partial_{A_2}\cw^{\prime\prime})-A_2^3\,\alpha\, \alpha^*=
 (\partial_{\alpha^*}\cw^{\prime\prime})\alpha^*-A_2(\partial_{A_2}\cw^{\prime\prime})=0.
\end{aligned}\end{align}

\end{document}